\documentclass[poms,nonblindrev]{poms1} % Use this for listing out authors' information submissions

\OneAndAHalfSpacedXI % % default line spacing !!!!!!!!!!!!!!!!!!!!!!!
\usepackage{graphicx}
\usepackage{subfigure, epsfig}
\usepackage{natbib}
\usepackage{caption}
\usepackage{wrapfig}
 \bibpunct[, ]{(}{)}{,}{a}{}{,}%
 \def\bibfont{\small}%
\TheoremsNumberedThrough     % Preferred (Theorem 1, Lemma 1, Theorem 2)
\ECRepeatTheorems

\EquationsNumberedThrough    % Default: (1), (2), ...
\begin{document}
%%%%%%%%%%%%%%%%

%\setcounter{page}{1}
%\DOI{10.1287/xxxx.0000.0000}%

\RUNAUTHOR{Acocella, Caplice, and Sheffi}

\RUNTITLE{Opportunities for market-based freight contracts}

\TITLE{The end of ``set it and forget it'' pricing? Opportunities for market-based freight contracts}

\ARTICLEAUTHORS{%
\AUTHOR{Angela Acocella, Chris Caplice, Yossi Sheffi}
\AFF{Center for Transportation \& Logistics, Massachusetts Institute of Technology,\\ \EMAIL{acocella@mit.edu, caplice@mit.edu, sheffi@mit.edu}} %, \URL{}}
} % end of the block

\ABSTRACT{
In the for-hire truckload market, firms often experience unexpected transportation cost increases due to contracted transportation service provider (carrier) load rejections. The dominant procurement strategy results in long-term, fixed-price contracts that become obsolete as transportation providers' networks change and freight markets fluctuate between times of over and under supply. We build behavioral models of the contracted carrier's load acceptance decision under two distinct freight market conditions based on empirical load transaction data. With the results, we quantify carriers' likelihood of sticking to the contract as their best known alternative priced load options increase and become more attractive; in other words, carriers' contract price stickiness. Finally, we explore carriers' contract price stickiness for different lane, freight, and carrier segments and offer insights for shippers to identify where they can expect to see substantial improvement in contracted carrier load acceptance as they consider alternative, market-based pricing strategies.
}

\KEYWORDS{Truckload transportation, freight procurement, supply contracts, indexed pricing} 
%Use this for final submission
%\HISTORY{This paper was first submitted on January 1, 2021 and has been with the authors for 3 months for 2 revisions.}

\maketitle
%%%%%%%%%%%%%%%%%%%%%%%%%%%%%%%%%%%%%%%%%%%%%%%%%%%%%%%%%%%%%%%%%%%%%%

% Samples of sectioning (and labeling) in POMS
% NOTE: (1) \section and \subsection do NOT end with a period
%       (2) \subsubsection and lower need end punctuation
%       (3) capitalization is as shown (title style).
%
%\section{Introduction.}\label{intro} %%1.
%\subsection{Duality and the Classical EOQ Problem.}\label{class-EOQ} %% 1.1.
%\subsection{Outline.}\label{outline1} %% 1.2.
%\subsubsection{Cyclic Schedules for the General Deterministic SMDP.}
%  \label{cyclic-schedules} %% 1.2.1
%\section{Problem Description.}\label{problemdescription} %% 2.

% Text of your paper here

\section{Introduction}
 
One of the unique aspects of the truckload (TL) freight industry is the non-binding nature of the contracts. Shippers (firms with goods that must be moved) contract with transportation service providers, or carriers (i.e., trucking companies) to haul their freight.

Contract prices are set at the time of the shipper's strategic transportation procurement event. The shipper communicates its forecasted demand for each lane (pickup origin to drop-off destination pair) with carriers upfront. However, volume and capacity commitments for both parties are not strict contractual obligations. First, the shipper is not required to offer the expected volume to the contracted carrier. In fact, the shipper may offer more than, less than, or even none of the awarded volume, with no direct financial penalty. This is because forecasting the precise time and lane of future demand for trucking capacity is not realistic, especially when projections are done often months or even a year in advance. Moreover, forecasting errors are amplified when shippers' end customer demand changes over the course of the transportation contract.

Second, contracted carriers are not obligated to accept 100\% of the volume that is offered or tendered to them. Responding to many shippers' uncertain demand makes committing a truck to a specific future time and location almost impossible, especially when commitments are expected to be in effect for a year or more in most cases. As a result, a carrier can enter a contract with a shipper for an expected amount of freight on a lane but uncertainties introduced from other customers, general market dynamics, or changes in the carrier's own business require the flexibility to reject loads in real-time capacity allocation decisions.

It is useful to think about the interactions between shippers and their for-hire motor carriers as two stages of TL transportation: 1) the strategic procurement process with its long-term carrier-lane matching decisions and 2) the shipper's operational, real-time load-carrier matching decisions with the corresponding carrier acceptance or rejection decisions. We describe these processes further in the following subsections.

\subsection{Strategic Transportation Procurement}\label{sec:FirstStage}
During the strategic procurement process (see \cite{Caplice2007} for a detailed description), shippers conduct a reverse auction in which they send a request for proposals (RFPs) to a group of carriers. The RFP documents the lanes (origin-destination pairs) on which the shipper requires priced transportation services. It also includes the expected volumes on those lanes and other service level expectations. Carriers respond to the RFPs by bidding the price they are willing to accept to serve the expected demand on each lane in which they are interested. The prices a carrier bids depend on the general attractiveness of serving the origin and destination regions, attributes of the freight itself, the carrier's previous experience working with that shipper, how well that lane fits within the carrier's existing networks, and the general market environment at the time.

The shipper awards the business to one or more carriers on each lane, which are referred to as the contracted, awarded, or primary carriers. Typically a carrier is selected because it is one of the lowest bidding carriers on the lane and has high expected level of service (e.g., load acceptance rates, on-time pickup and delivery, technological sophistication for communication, information and data exchange, and payments, etc.).

Due to the possibility that primary carriers may reject loads, shippers construct a routing guide for each lane. The routing guide is a sequential list beginning with the primary carriers followed by a set of backup carriers. These backup carriers typically have unsuccessfully bid on the lane, thus expressing ability and willingness to serve at least some of the demand on that lane. Most often, these carriers had bid higher prices than the winning carrier(s). The backup carriers are entered into the routing guide at their bid price, however, there is no contract in place between the shipper and backup carriers. This means the backup carriers' routing guide price is not binding. Further, there is a much lower implicit assumption that backup carriers will produce capacity.

\subsection{Operational Load Tendering, Acceptance \& Rejection}\label{sec:tenderacceptintro}
At the time a load needs to be moved, the shipper's transportation management system (TMS) tenders the load to the first carrier in the routing guide (the primary, awarded carrier) at the contracted price. If the carrier accepts the load, it agrees to move it at the contracted price. If the primary carrier rejects the load, the load is offered to the first backup carrier at the price denoted in the routing guide. Not only is this price between shipper and backup carrier non-binding, it is likely higher than the primary carrier's price. Depending on prevailing market conditions, shippers price escalations with backup carriers reach more than 18\% above the primary carrier's contracted price \citep{Acocella2020a}.

Like primary carriers, backup carriers may accept or reject load offers. The shipper proceeds down the routing guide until a backup carrier accepts the load or some price or time threshold is hit. At that point, the shipper may turn to the spot market. In this case, the shipper consults a load board where it can post information regarding the load's pickup, drop-off, timing, and other requirements and view carriers who have posted currently available capacity and associated prices on the lane.

\begin{wrapfigure}[13]{l}{0.6\textwidth}
  \begin{center}
  \vspace{-\intextsep}
    \caption{Daily Spot Prices for Single Lane}
    \label{fig:DailySpotPrices}
    \includegraphics[width=0.58\textwidth]{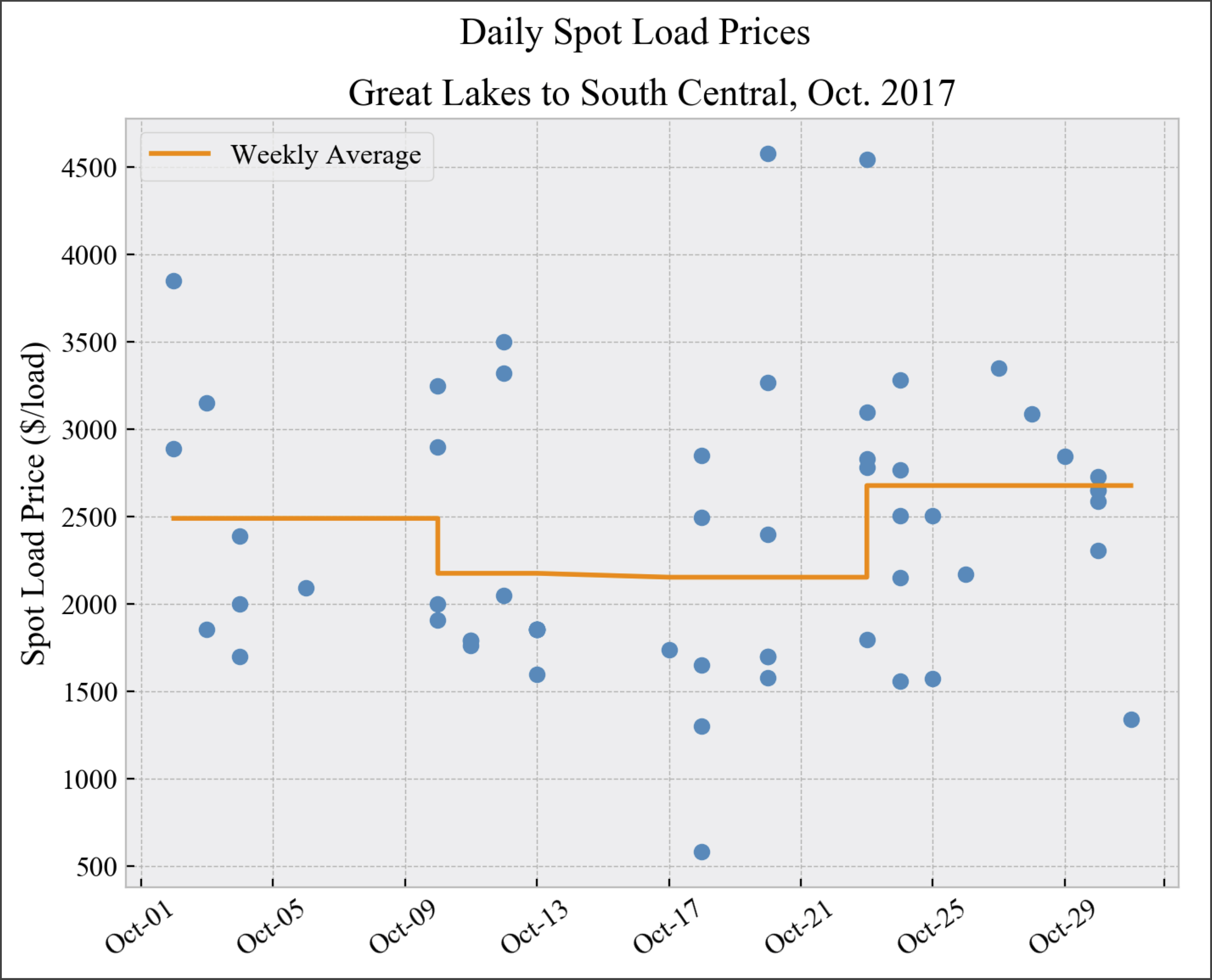}
  \end{center}
\end{wrapfigure}

Shippers find a carrier on the spot market for independent transactions, but there is no single spot market price at any given time, even for spot loads on the same lane. Instead, when we refer to the ``spot market price'' it is the average of a range of prices. Figure \ref{fig:DailySpotPrices} illustrates this point.  It depicts actual loads fulfilled on the spot market from 6 shippers on a single lane in October 2017. Each day, there may be one, many, or no loads that require spot capacity. Each load is fulfilled at a different price set by the individual carrier and depends on the carrier's network structure at that moment. In the remainder of this study, we discuss the dynamic average lane-specific spot price, depicted in the figure, which represents an underlying distribution of individual realized spot load prices.

\subsubsection{Contractual relationships and the best known alternative}
Spot market prices are highly volatile and represent the dynamic nature of immediate balance between trucking supply and demand. For shippers and carriers alike, the spot market represents alternative price options to a contracted load. When trending spot prices are high relative to contract prices, carriers may be tempted away from adhering to their contractual agreements. A primary carrier can either uphold its contractual commitments with the shipper by accepting loads and maintaining the relationship, or it can act opportunistically by rejecting loads and offering its capacity on the spot market with higher expected profit. However, by reneging on its contract and rejecting loads, the carrier risks potential future business from that shipper.

% It is important to note that while higher external (spot) market prices may incentivize carriers to act opportunistically and reject contracted volume for spot loads \cite{Williamson1975}, contract carrier load rejections are not necessarily demonstrating opportunistic behavior. However, when the expected profit \cite{Scott2020}

% ADD transition sentence or move this somewhere else
About 72\% of for-hire TL freight is accepted by the primary carrier, while about 5\% is fulfilled by spot capacity \citep{Caplice2007}. The remaining 23\% is moved by backup carriers within the shipper's routing guide. When spot market prices are higher than average contract prices - a condition referred to as a tight market - shippers typically pay a price premium of 35\% above the contract price if they end up using spot carriers \citep{Yuan2019}.

Even when backup carrier prices or spot market prices are below a shipper's contracted carrier's price, there are substantial benefits to working with primary carriers rather than backup or spot carriers. The shipper's RFP is a vetting process. Over the course of weeks, sometimes months, the shipper and its bidding carriers not only communicate expected demand, service expectations, and pricing, but the carriers have also have demonstrated fast and simple communication processes, and technological sophistication to integrate with the shipper's electronic data interchange (EDI) and payment systems.

An account manager is assigned by each of the contracting parties. This person continuously evaluates performance over the course of the contract and is the point of contact when performance issues arise. As a result, the value for the shipper of working with a primary carrier includes the ease of business processes and relationship developed. Due to its transactional nature, none of these process or communication channels are established for spot loads. Furthermore, the spot market price is not known to the shipper in advance. Thus, both service level and actual price to be paid to spot market carriers are largely uncertain. Therefore, it is typically in shippers' best interest to encourage primary carriers to accept loads, even if the spot market price is expected to be better (i.e., lower) than contract prices.

As freight markets fluctuate between periods of over and under supply - soft and tight markets, respectively - shippers' fixed contract prices can become stale. As a result, the amount of freight that is accepted and moved at the original contracted price declines over the course of the contract. To mitigate this issue and ensure contract prices remain market competitive, there has been growing interest from both shippers and carriers to explore market-based pricing into their portfolio of TL freight contracts. Understanding carriers' willingness to stick to (or defect from) their contracted load price helps practitioners identify the most promising network, lane, freight, and carrier segments for this market-driven approach.

The remainder of this paper is structured as follows. Section \ref{sec:indexedcontracts} discusses practitioners' considerations regarding index-based contracts, Section \ref{sec:litreview} offers a review of the relevant literature, Section \ref{sec:RQ} summarizes our research question and the hypotheses tested, and Section \ref{sec:acceptmodels} describes our empirical dataset and model specifications. We present the results of the models in Section \ref{sec:results} and their implications, particularly for market-based pricing consideration, in Section \ref{sec:implictions}. We conclude in Section \ref{sec:futureresearch} with a discussion of limitations of this research and areas for further exploration.

\section{Market-based Freight Contracts} \label{sec:indexedcontracts}
For every load a contracted carrier is tendered on a contracted lane, it can either accept the load or reject it and fill available capacity with a load on the spot market. Carriers may also have other contracted shippers on the same or similar lanes and thus other contracted business available as an alternative option. However, because the shipper has no knowledge of its contracted carriers' other customers and because spot prices and contract prices move up and down together, albeit with some lag (see \cite{Coyote2018}), we use the average lane-specific spot market price to represent carriers' alternative options.

As the average price of spot market loads available to the carrier increases relative to the contracted price, an opportunistic carrier with low contract price stickiness will be incentivized away from accepting the contracted shippers' loads. For each load offered to a contracted carrier, we calculate its Spot Rate Differential (SRD), or how much the current lane-specific spot price is above or below the load's contract price, as a percentage:

\begin{equation} \label{eq:SRD}
SRD_{k,i,j,t} = \dfrac{(SpotPrice_{i,j,t} - ContractPrice_{k,i,j,t})}{ContractPrice_{k,i,j,t}} \times 100
\end{equation}

where $k$ is the load tendered to the carrier on lane with origin $i$ and destination $j$ at time $t$.

This Spot Rate Differential is the percent difference between the contract price and spot market price at a given time and the key metric for carrier contract price stickiness. Building off of the literature described in Subsection \ref{sec:litpricesens} we define contract price stickiness as the rate at which the contract price must change (relative to the current lane-specific spot price) as factors exogenous to the contract change.

A market-based pricing strategy aims to appeal to carriers with low contract price stickiness. Market- or index-based pricing is commonly used in industrial, agricultural, and energy commodities markets. Similar to the truckload market, these products see cyclical supply and demand fluctuations. Market-based pricing helps simplify price negotiations and increase transparency for long-term contracts between sellers and buyers, particularly when price volatility over time is a concern \citep{Deloitte2016}. Interest in indexed pricing has grown in the truckload freight market as both shippers and carriers seek ways to mitigate the risks incurred by freight market cycles.

\begin{wrapfigure}[13]{r}{0.6\textwidth}
  \begin{center}
  \vspace{-\intextsep}
    \caption{Industry Index, Spot, and Contract prices}
    \label{fig:IndexLoadPrice}
    \includegraphics[width=0.58\textwidth]{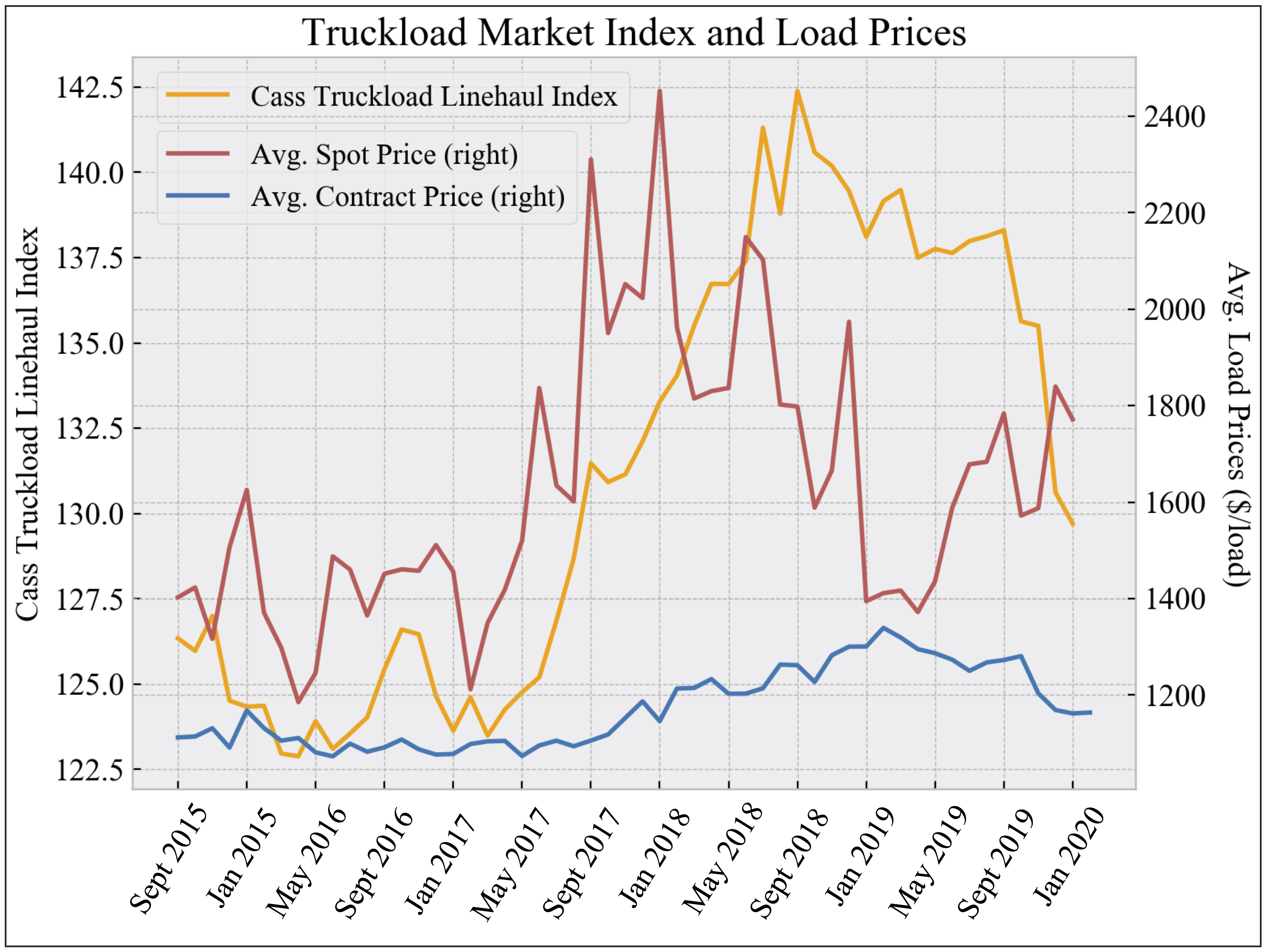}
  \end{center}
\end{wrapfigure}

Figure \ref{fig:IndexLoadPrice} demonstrates the dynamic nature of the truckload freight market. We present the Truckload Linehaul Index reported by Cass Information Systems, Inc. - a leading industry provider of information and payment solutions - and national average contract and spot market prices from our empirical dataset from September 2015 to January 2020.\footnote{The Cass Truckload Linehaul Index is comprised of 95\% contract load transaction data and 5\% spot load data. Discussion of how our empirical dataset appropriately represents dynamics in the truckload freight market is found in Section \ref{sec:acceptmodels}}.  
\indent The variations in market index and spot and contract prices over this time period reflect both soft and tight market periods. Until July of 2017, the industry experienced a soft market, where we see low prices and index values. Following this time, the industry experienced a very tight market, with high prices and severely deteriorated primary carrier acceptance rates. In fact, in the first soft market period, average primary carrier acceptance rates were 81.9\%. This number dropped to 68.5\% in the subsequent tight market period. \cite{Acocella2020a} provide quantitative justification of these market periods and, along with \cite{Yuan2019} and \cite{Sokoloff2020}, analyze market-driven primary carrier acceptance rates.

As the freight market cycles between soft and tight periods, index-based pricing strategies aim to allow contract prices to adjust automatically with changes in the broad market. Otherwise, shippers manually adjust prices through full RFPs or ``mini-bids'' (targeted carrier or lane price adjustments specifically focused on under-performing segments, typically resulting in short-term fixed-price agreements). With index-based strategies, the contracted price between a shipper and primary carrier increases or decreases with a market-based indicator such as those reported by  Cass Information Systems Inc. or DAT Freight \& Analytics.

A handful of large shippers that operate in the consumer packaged goods, food and beverage, and manufacturing industries, and their carriers (both asset and non-asset) have explored index-based pricing strategies in practice. Those that have to date, however, have only done so on a small set of trial or pilot lanes \citep{Sinha2019}. In addition, some carriers, particularly 3PLs and brokerages (i.e., non-asset providers), have dedicated resources to offer dynamic pricing options for their shipper customers (see \cite{Convoy2020}, \cite{Schneider2019}, \cite{Uber2021}, and \cite{BRWilliams2020}).

Despite the potential benefits of market-based pricing strategies, the advantages of traditional fixed-price contracts remain. Shippers and carriers alike seek consistency and predictability - both in terms of demand or capacity availability and prices. Shippers run annual RFPs to establish fixed contract prices around which they can budget. Carriers intend to provide high-quality service to their customers but need to know they will be able to cover their costs in the process. Keeping this in mind, we aim to help shippers and carriers understand where this traditional fixed-price approach is effective and where an alternative market-driven approach may be beneficial.

\section{Literature Review} \label{sec:litreview}
In this research, we quantify carrier's willingness to stick to contracts by constructing empirical behavioral models that include industry dynamics not captured in previous literature.

\subsection{Supply chain contracts} \label{sec:contracts}
Much of the extant literature on supply chain contracting explores the ways contracts help coordinate or share risks between buyer and supplier \citep{Lariviere1999}. Agents seek risk-sharing contracts to encourage both sides to remain committed to the contract terms through buy-back \citep{Pasternack1985, Deneckere1997}, revenue sharing \citep{Cachon2005}, and options \citep{Barnes2002} contracts. (See \cite{Cachon2003} and \cite{SimchiLevi2014} for surveys of these contracting mechanisms.)

The non-binding TL freight contract is similar to the self-enforcing agreements studied in the context of repeated games literature, which explores agreements that result in both parties maintaining the terms of the agreement over time without the interference of an external party \citep{Telser1980}. Instead, informal mechanisms are used for contract enforcement. As the transportation provider considers each load, if it sticks to the contract by accepting the load, it maintains that relationship with its shipper. On the other hand, the transportation provider can defect from the contract, reject the load, and damage the relationship with each rejection (see \cite{Scott2020}). This self-interested, opportunistic behavior that violates the existing contract may be influenced to higher priced alternative options on an external spot market \citep{Williamson1975, Williamson1985, Wathne2000}. However, the value of future potential business - i.e., the shadow of the future - encourages carrier contract compliance \citep{Heide1992, Lumineau2012}.

For shippers, the incentive to stick to their contracts stems from the initial investment and vetting involved with securing the contracted carriers in the first place. The procurement process discussed in Section \ref{sec:FirstStage} takes months of preparation, communication, and analysis to establish contracted carriers on each lane. Transaction Cost Theory considers the costs incurred as a result of the information search, negotiation, and monitoring involved in the exchange of a good or service between agents – effectively, the frictions of a transaction \citep{Williamson1985, Williamson1979}. The theory justifies the use of contracts as a means of defining the terms of inter-firm agreements \citep{Williamson2002}, made particularly important with uncertainty and high frequency of interactions \citep{Williamson1979}. \cite{Masten2009} underscores the benefit of contract use, noting that the contract collects many repeated heterogeneous transactions under one common pricing policy, reducing the transaction costs needed to repeatedly negotiate prices for each transaction.

\subsection{Shipper-carrier relationships} \label{sec:relationships}
We further motivate the research by expanding the extant body of literature on shipper-carrier relationships, specifically that on primary carrier load acceptance decisions. Rather than direct measures of the shipper-carrier interactions, however, much of the literature considers attributes of the lanes and freight to determine primary carrier acceptance rate (PAR). PAR is measured as the percentage of loads a primary carrier accepts relative to the number of loads it is tendered on its contracted lanes. High lane volume \citep{Harding2005}, low lane volume volatility \citep{Kim2013}, high prices \citep{Amiryan2015}, and high lane consistency, or cadence, \citep{Yuan2019} have been found to be positively correlated with higher PAR, all else equal.

The shipper-carrier relationship itself, however, has been found to contribute to carrier PAR in a few studies. \cite{Scott2016} analyze contract and spot market transactions and the impact of the history of the shipper-carrier relationship on freight acceptance. The authors find that less frequent load offers increase the likelihood of a primary carrier rejecting a load, while higher offered volume, lower load offer volatility, and higher revenue transacted between the shipper and carrier increase the likelihood of carrier's load acceptance. \cite{Zsidisin2007} define a ``good'' shipper-carrier relationship as one in which a contract is in place and find that contracted carriers outperform non-contracted carriers in terms of freight acceptance, on-time delivery, and pre-positioned capacity. These two studies each obtain data from a single shipper, which limits their ability to generalize across types of shippers or segment their datasets to offer insights specific for types of shippers' business.

The impact of market dynamics on carriers' freight acceptance decision is studied in \cite{Acocella2020a}. The authors consider two distinct market conditions: soft, where shippers have low demand relative to available capacity and thus corresponding low contract and spot prices, and tight, where demand for TL capacity outstrips supply and prices are high (spot prices typically exceed contract prices). The authors find one of the main contributor to primary carriers' tight market acceptance decision to be how competitive the shippers' contracted load offer price is with prevailing market prices. \cite{Scott2016} also consider a measure of market condition - Spot Premium - in determining carrier load acceptance decisions and find that carriers consider overall market conditions and alternative priced loads when making contracted load acceptance decisions.

The nature of existing TL contracts, external market conditions, and their impact on carriers' load acceptance decisions are studied in \cite{Scott2020}. The authors find that explicit contracts (those that have more formally defined service level expectations) as compared to implicit contracts illicit higher primary carrier acceptance rates from carriers. In addition, and similar to the findings by \cite{Acocella2020a}, as the market tightens and becomes more attractive to carriers, the benefit seen by these explicit contracts diminishes.

Finally, \cite{Lindsey2015b} consider TL carriers' reservation prices and willingness to accept loads. Using spot market prices and a discrete choice survey experiment, the authors find that carriers' contracted load acceptance decisions are impacted by higher-priced hypothetical alternative load options. We expand this research by using empirical data to study actual carrier decisions and explore how these decisions differ across shippers' freight networks, carriers' service types, and overall freight market conditions.

\subsection{Supplier price stickiness} \label{sec:litpricesens}
While our research builds on the transportation literature, we also draw from that which models price stickiness - in other words, the rate at which prices change in response to internal firm or external market dynamics. Much of this literature focuses on how suppliers' prices change as a collective at the industry level, rather than the micro, firm level.

The most relevant stream of literature to our research consists of a set of studies focused on producer pricing stickiness to customer demand patterns or to external market prices. Manufacturers adjust prices due to underlying costs and customer demand changes \citep{Loupias2013} and beliefs that competitors are also changing prices (known as coordination failure) (\cite{Blinder1991} and \cite{Blinder1998}).

This coordination failure comes into play in the TL context during the strategic RFP stage. Carriers do not coordinate with one another on pricing of lanes. However, while carriers submit lane bid prices according to their own network fit, if they want to be competitive, they must also factor in their beliefs about how other carriers may be pricing the lanes they want. Submitting bids that are too high may mean another carrier is awarded the business; too low and they may be chosen for the contract, but at a price that cuts into critical profit margins. This coordination failure may result in a race to the bottom and cause the shipper to award poor-fitting carriers to lanes.

Studies by \cite{Fabiani2006} and \cite{Fabiani2007} use a survey of over 11,000 firms across Europe and find that most firms consider both historical market information and expectations of the market when making pricing decisions. The authors conclude that not only is price stickiness influenced by market conditions, it is also related to customer relationships as defined by explicit and implicit contracts. These findings underscore truckload practitioner sentiments and findings by \cite{Scott2020} that a combination of market conditions and standing shipper-carrier relationships are key determinants for carrier pricing and load acceptance decisions.

The above studies also find that individual firms' rate of price change tends to be slower than that of the general market. This time- and market state-dependent supplier price stickiness is further explored by \cite{Loupias2006}, \cite{Hall2000}, and \cite{Apel2005}. Similar results are observed in the TL freight context as carriers' contract price changes tend to lag those of spot prices (see \cite{Coyote2018}). In addition, contract freight rates rise more quickly than they fall.

The extant literature shines light on how producers make (predominantly strategic) pricing decisions. However, there are few studies of if or how suppliers choose to offer their product or service at an operational level to contracted customers or to non-contracted customers on a spot-like market. We aim to add to the literature in this way.

\section{Research Question and Hypotheses} \label{sec:RQ}
As discussed above, the long-term, fixed-price contracts between shippers and their TL carriers are non-binding in volume and capacity commitments. They often result in degraded performance and unexpected price escalations as exogenous market conditions change. Alternative contract forms, in particular those based on indexed pricing, have drawn recent attention. With this research, we aim to help shippers and carriers determine the most promising areas within their networks for index-based pricing in their portfolio of freight contracts. Thus, we address the following research question:\\
\textbf{For which segments of shippers' networks - lanes, volume, demand patterns, and carriers - should they consider index-based contracts?}

To answer this question, we model primary carrier load acceptance decisions based on the load's offered price relative to the carrier's best-known alternative option: the current lane-specific spot market price. In this way, we can demonstrate carriers' contract price stickiness. We formulate our first hypothesis as:

\textit{H1: A load is more likely to be accepted by a primary carrier the higher the contracted price is relative to the current, lane-specific spot market price.}

This claim is the basis for our behavioral assumptions that carriers prefer higher priced loads \citep{Williamson1975, Wathne2000}.

Next, we consider carriers' willingness to stick with contract prices for different lane and freight types. We measure how primary carriers' likelihood of accepting a load is impacted as the offered price of the load changes relative to the spot market price for these lane and freight segments of interest. First, we consider a lane's distance. While carriers each have individual preferences and strategies relating lane distances, we aim to isolate segments to determine general patterns. We define the following hypothesis:

\textit{H2: The likelihood a primary carrier accepts a load increases with lane distance and further increases as the load's contracted price increases relative to the current, lane-specific spot price.}

Next, we consider how price adjustments impact primary carrier load acceptance for different lane demand patterns. Shippers' infrequent and inconsistent tendering behaviors lead to lower primary carrier acceptance (see Section \ref{sec:relationships}). For lanes on which a shipper tenders loads inconsistently, it may be more difficult or require longer distances driven empty for a carrier to re-position capacity when a load is tendered. A higher price incentive may improve the probability the carrier accepts the load.

We expect that index-based pricing for lanes with inconsistent demand will result in higher primary carrier acceptance and less reliance on backup or spot alternatives. Moreover, carriers would benefit by being able to better cover additional internal costs for serving unanticipated demand. This leads us to the following two hypotheses:

\textit{H3: The likelihood a primary carrier accepts a load increases with higher shipper tendering frequency on the awarded lane and further increases as the load's contracted price increases relative to the current, lane-specific spot price.}

\textit{H4: The likelihood a primary carrier accepts a load increases with lower shipper tendered volume volatility on the awarded lane and further increases as the load's contracted price increases relative to the current, lane-specific spot price.}

In the above hypotheses, we capture lane-level demand patterns that make carrier acceptance difficult. Next, we measure a load characteristic: surge volume. Recall, the contract agreement between a shipper and a primary carrier includes a (non-binding) awarded or expected volume However, due to poor forecasts, network changes, or other unanticipated end customer demand, shippers may tender more loads to a contracted carrier than the awarded volume in a given week.\footnote{A similar issue for primary carriers may arise if the shipper tenders much \textit{less} than the awarded volume. While this is an important issue to address, it is out of the scope of the present research.} This additional surge volume above the carrier's allocated capacity to that lane may be more difficult or costly to cover and may require an additional price incentive for the carrier to accept the surge loads. Thus, we identify each load that is tendered above the awarded weekly volume as surge loads and test the following hypothesis:

\textit{H5: The likelihood a primary carrier accepts a surge load increases as the load's contracted price increases relative to the current lane-specific spot price.}

Finally, we aim to determine which carrier types are best suited for indexed pricing. Carriers fall into one of two categories based on the services provided: asset-based carriers that own the trucks and trailers used to move freight, and non-asset providers, often referred to as brokers or third party logistics (3PL) providers. This latter group of providers remove the shippers' burden of securing capacity and act as the middle man between shippers and asset-based carriers. Brokers access a vast pool of typically smaller, asset-based carriers, aggregate their capacity, and match it to shippers' demand. These brokers, or non-asset providers, typically buy and sell transportation based on their expectations of the general market; their profit margins are tied to how well they are able to manage the cycles. We test the following hypothesis:

\textit{H6: Non-asset primary carriers are less likely than asset-based primary carriers to stick to fixed-price contracts as spot market prices rise.}

We test this hypothesis by calculating the rate of primary carrier load acceptance (PAR) as Spot Rate Differential changes. A higher rate of change (i.e., decrease in PAR as spot price increases relative to contract price) suggests lower contract price stickiness. A similar approach is taken by \cite{Lindsey2015b} to measure carriers' load price elasticity.

We further characterize asset carriers based on their fleet size. We expect that larger carriers may be better insulated from fluctuations in the market and may be more willing to forego opportunistic spot market loads than smaller carriers. We formulate our final hypothesis as follows:

\textit{H7: Larger asset carriers are more likely than smaller carriers to stick to fixed-price contracts as spot market prices rise.}

By exploring each of these hypotheses individually, we can measure carriers' contract price stickiness for different freight segments of interest and identify which areas of a shipper's network each carrier type may be most amenable to index-based pricing. In this way, we address our central research question.

\section{Carrier load acceptance model specification} \label{sec:acceptmodels}
In this section, we summarize our empirical data, describe our carrier acceptance model, and define the variables we use to build the model. Our partner company - a major US freight transportation management firm - provides us with transaction data over four years (2015-2019) of all of the TL loads for 68 shippers of various sizes and industry verticals and their 412 (primary and backup) carriers, both asset and non-asset. The data represent each load's tender sequence. This includes the date, time, and price at which each load is tendered to the primary carrier with its accept or reject decision, and, if needed, backup carriers' accept or reject decisions. The reported tender sequence continues down the routing guide waterfall until a carrier - primary or otherwise - accepts the load and the price the load is accepted at, including an indication if it ultimately is moved on the spot market.

The set contains 1.7 million long-haul loads (i.e., loads that move a distance greater than 250 miles)\footnote{We use this long-haul distinction because pricing structures for the alternative, short-haul moves, differ from those we consider and discuss in this research.} all of which originate and terminate in the continental US. From the subset of tenders to primary carriers, we model the probability a load is either accepted or rejected (a binary outcome) by the primary carrier based on associated load, lane, shipper, and carrier characteristics described in the following subsections.

Logistic regression models are widely used in econometric literature to isolate the relationships between a binary dependent variable and independent input variables. Moreover, these models are the predominant modeling choices for authors studying producer and consumer price stickiness \citep{Loupias2013, Cecchetti1986, Cao2012}. We adopt a logistic model choice as well and, to allow for non-linearity between independent and dependent variables \citep{Morris1988}, we discretize continuous input variables and include them as categorical variables.

Our dataset is comprised of multiple load accept/reject decisions by each carrier. As such, we must account for repeated measures of the same individual carrier. With repeated measures data, typical (logistic) regression modeling neglects to account for the correlations between the set of decisions made by the same individual. This within-subject correlation results in inefficient estimators. In other words, the calculated estimators have a greater spread around the true population values. Instead, General Estimating Equations (GEEs) model the average response of an individual in the population \citep{Liang1986, Ballinger2004}. The regression coefficient estimates in GEE models consider the covariance matrix between the outcomes in the sample associated with the same individual. GEE estimators reduce to those obtained through OLS if the dependent variable is normally distributed and no within-individual response correlations exists \citep{Hardin2012, Greene2003}.

We use a GEE model with a logistic link function where the coefficient estimates represent the marginal increase or decrease in the (log) odds a carrier accepts a load:

\begin{equation} \label{eq:logit}
    logit(y_{c,k}) = log\Big(\dfrac{y_{c,k}}{1-y_{c,k}}\Big)=x_{c,k}^T\boldsymbol{\beta}
\end{equation}

where each individual primary carrier, $c$, makes a binary accept/reject decision for each load it is tendered, $k$ (the positive outcome, $y_{c,k}=1$, is a load acceptance), and the matrix of explanatory variables, $x_{c,k}$, are the lane, freight, shipper, and carrier variables described in the following subsections.

\subsection{Load Spot Rate Differential} \label{sec:SRD}
We define the Spot Rate Differential (SRD) in Section \ref{sec:indexedcontracts} and Eq. \ref{eq:SRD} as the percent difference between the current lane-specific spot market price and the contracted price of a load. The SRD calculation requires knowledge of each lane's spot price at the time a load is offered. While we do not have such data consistently across all lanes and time, given the breadth of our dataset, we can calculate benchmark spot market prices.

Our dataset approximately represents the general freight market trends as it is comprised of many shippers' load tenders across the continental US. We corroborate this claim by comparing two statistics from our dataset to external industry data. First, we measure the correlation between the time series of average primary carrier acceptance rate (PAR) in our dataset, a real-time primary carrier behavior, and the Morgan Stanley Freight Index, which represents overall practitioner sentiment of the market's supply and demand. The correlation between the two time series is 85.2\% (see \cite{Scott2016} for a similar justification process). Second, the correlation between our national average contract linehaul price and that of the Cass Truckload Linehaul Index is 91.8\%. We conclude that our dataset is sufficiently representative of the overall freight market.

Next, we reconstruct spot prices for each origin region to destination region combination. Regions of the US differ in attractiveness to carriers based on the business opportunities that are expected in those areas. For example, regions with high outbound demand are typically more attractive for carriers to accept loads going into because they are more likely to easily find a follow-on load. To account for the dynamic nature of spot prices, we include the year and month indicator in the spot price model.

\cite{Daganzo2004} shows that point-to-point transportation costs (e.g., linehual costs) result from a combination of fixed and variable costs. Based on this, we model spot prices using multiple linear regression with heteroskedastic robust standard errors. We regress the point-to-point linehaul price of loads that are accepted on the spot market in our dataset on origin and destination region binary variables and a month and a year binary variable. These represent fixed costs. We include a continuous distance variable, which represents the carriers' variable costs \citep{Ballou1991,Scott2015}. \cite{Acocella2020a} further detail the price benchmarking methodology used here.

Our base case lane (i.e., the binary variables omitted to avoid multicollinearity) originates in the Lower Atlantic region of the US and terminates in the South Central region in January of 2016. These ommitted variable choices correspond to the most volume (i.e., greatest number of observations) within each categorical variable. The lane-specific spot price for a given time, $Spot_{i,j,t}$ is defined as:

\begin{equation} \label{eq:Spot}
\begin{split}
Spot_{i,j,t} &= \hat{\beta}_{base} + \hat{\beta}_{dist} \overline{X}^{(i,j)}_{dist}  + \sum_{i\in I, i\neq i_{base}} \hat{\beta}_i X_i + \sum_{j\in J, j\neq j_{base}}\hat{\beta}_j X_j\\
& + \sum_{m\in M, m\neq m_{base}}\hat{\beta}_m X_m + \sum_{y\in Y, y\neq y_{base}}\hat{\beta}_y X_y
\end{split}
\end{equation}

The intercept term, $\hat{\beta}_{base}$, is the fixed cost of the base case, and $\hat{\beta}_{dist}$, is a distance (i.e. variable cost) coefficient associated with the average distance of loads between $i$ and $j$, $\overline{X}^{(i,j)}_{dist}$. The fixed cost of the origin and destination regions that are different from the base case are represented by the $I$-1 origin coefficients, $\hat{\beta}_i$ (where $I$ is the set of origin regions and $X_i$ the corresponding $I$ binary variables indicating in which region the load originates), and $J$-1 destination coefficients, $\hat{\beta}_j$ (where $J$ is the set of destination regions and $X_j$ the corresponding destination binary variables). The 15 mutually exclusive and collectively exhaustive regions are key market areas defined by our industry partner representing geographic clusters of transportation demand patterns. The origin and destination coefficients of our linear regression model, $\hat{\beta}_i$ and $\hat{\beta}_j$, can be interpreted as spot price premiums associated with an origin or destination different from the base case lane.

Finally, $\hat{\beta}_m$ and $\hat{\beta}_y$ measure the dynamic, time-based changes in spot prices for each month and year, respectively, and capture both seasonal and underlying market structural trends.\footnote{We choose to include month and year variables (as opposed to a single variable that treats each month in the time frame separately - that is, $m \in \{1,60\}$ and no $X_y$ variable) to ensure enough observations are in each time-based outcome variable for robust coefficient estimates.} As a notational simplification, for the remainder of this paper, we combine $m$ and $y$ for a time-dependent variable by denoting it with a subscript $t$.

With the results of this model, we calculate a single average spot price for every origin-destination-month-year combination in the dataset. As discussed in Section \ref{sec:tenderacceptintro}, this spot price represents the average of a distribution of underlying spot load prices. We use this average spot market price to calculate the Spot Rate Differential (SRD) for each load, $k$, using Equation \ref{eq:SRD} as the key metric to test carriers' contract price stickiness.

\subsection{Lane distance} \label{sec:dist}
Carriers' preferences regarding lane distance may vary depending on individual operations. Each carrier tries to maximize asset utilization - in other words, miles driven carrying a paid load. We measure a lane's distance by the number of days it takes to drive that distance, assuming a driver can drive 400 miles per day based on federally mandated hours of service regulations and actual driving patterns \citep{JBHunt}. For a more detailed discussion of carrier economics, see \cite{Belzer2018}, \cite{Masten2009}, and \cite{Burks2019}.

Our lane distance measure is broken into five binary ``travel-days'' variables based on whether the lane is expected to take up to 1 day to drive, 1-2 days, 2-3 days, 3-4 days, or more than 4 days. Rather than using a continuous variable for distance (or travel-days), we decompose the variable into these categories to avoid a linear assumption on the relationship between carrier acceptance and distance.

\subsection{Lane demand cadence} \label{sec:cadence}
In the supplier-buyer relationship, frequency of interactions points to more positive relationship; infrequent demand patterns are problematic for suppliers \citep{Rinehart2004}. This aspect of the relationship is particularly true in the TL industry. Inconsistent or infrequent tendered volume from one shipper (the customer) makes it difficult for the transportation suppliers (carriers) to plan where and when they need to position capacity to balance their networks and serve each of their other customers.

One metric carriers use to measure shipper performance is tender cadence, or the frequency at which loads are tendered \citep{JBHunt, CHR2015}. \cite{Scott2016} incorporate this frequency of interactions as a contributor to load acceptance by measuring the number of days since the previous load was offered to a carrier by a shipper. The authors find that carriers are less willing to accept loads that are tendered at unpredictable frequencies. Moreover, \cite{Acocella2020a} demonstrate that primary carrier acceptance increases with the percentage of weeks in which loads are tendered to that carrier on that lane. 

We use this latter measure to characterize a lane's frequency. For each load, we calculate the percent of the preceding four weeks in which that shipper tendered at least one load to that carrier on that lane. The resulting cadence metric is a discrete measure, taking on values of 0\%, 25\%, 50\%, 75\%, or 100\%. Our input variables to the carrier acceptance decision models are binary variables indicating which of these discrete values the cadence measure takes.

\subsection{Lane demand volatility} \label{sec:volatility}
In addition to the frequency of interactions, consistency of demand is an important factor for carriers to anticipate capacity needs \citep{JBHunt, CHR2015}. Both practitioners and the literature note the importance of reducing tendered volume variability to improve carrier freight acceptance, reduce the cost of loads, and help carriers better utilize their capacity \citep{Harding2005, Scott2016, Yuan2019, Acocella2020a}.

We incorporate a measure of lane-level tendering volatility between a shipper and primary carrier acceptance. Each load is assigned the corresponding lane-level tendering volatility, which we calculate as the shifted 4-week rolling average week-over-week change (measured as the square difference) in tendered volume from the shipper to the carrier on that lane:

\begin{equation} \label{eq:Vol}
    Vol_{i,j,t} = \sqrt{\dfrac{\sum^{t-4}_{\tau=t-1}(d_{i,j,\tau} - d_{i,j,\tau-1})^2}{4}}
\end{equation}

where load $k$ is tendered on lane $(i,j)$ in time period $t$ and $d_{i,j,\tau}$ is the number of loads tendered on the lane in week $\tau$. We consider only weeks in which loads materialize and are tendered to the carrier - in other words, weeks in which $d_{i,j,\tau} > 0$. This is because we want a measure of the volatility of the \textit{materialized} volume. We already capture how frequently there are no-volume weeks with the Cadence measure.

\subsection{Surge volume} \label{sec:surge}
Next, we consider surge volume, a load characteristic defined as tendered loads that are above the awarded, or expected, weekly volume. Carriers report that they can typically manage to make capacity available when the number of loads tendered from a shipper in a week is within about 10\% of the lane award volume. However, as volume reaches and surpasses 20\% of the award volume, carriers often are unable to serve the excess demand. Moreover, shippers commonly call for contracted carriers to flex up with increased demand, often up to 20\% above the awarded volume, as a stipulation of the service level expectations in the contract \citep{Singh2021}. Such service level expectations highlight the non-binding nature of the contract: while agreed upon and defined in the contract, they are not legally or contractually enforceable by the shipper. The main incentive for the carrier to uphold them is the promise of continued business from the shipper (i.e., the shadow of the future).

While the expected volume is part of the information communicated to carriers during the RFP, many shippers do not keep careful record of the awarded volume to each carrier on each lane after the bid is complete. As such, our dataset does not include the primary carriers' awarded volume for each lane. Instead, we use a proxy for this awarded volume: the preceding shifted 4-week rolling average of the tendered volume to the primary carrier on the lane. \cite{Scott2016} use a similar proxy to rank loads. The authors measure the average daily volume on a given lane over the 30 days preceding the load of interest and denote how the load's rank within the day measures relative to that 4-week rolling average daily volume. We similarly rank each load within the week and categorize it based on how its rank compares to the awarded weekly volume proxy.

Each load is assigned to its corresponding surge category. If the load's rank within the week is less than or equal to the awarded volume proxy, it is given a Surge category of \textit{Within Mean}. If the load's rank is more than the average but less than or equal to 10\% above the mean, it is in the \textit{Up to 10\% Surge} category, if it is more than 10\% but less than or equal to 20\% of the average volume, it is in the \textit{Up to 20\% Surge} category, and if the load's rank is more than 20\% above the awarded volume proxy, it is in the \textit{Over 20\% Surge} category. We expect the higher the surge volume category a load is in, the less likely it is to be accepted by a primary carrier.

\subsection{Carrier service type}\label{sec:servicetype}
Asset and non-asset carriers offer different services which drive their relationships with shippers and how they interact with the overall market. On one hand, asset carriers own the trucks (tractors) they operate. They have fixed available capacity to manage across their networks and serve customers.

On the other hand, non-asset carriers - otherwise known as brokers or freight forwarders - do not own the tractors or trailers that move their customers' loads. Instead, non-asset providers match shippers with asset-based carriers. The advantage to shippers of working with a non-asset carrier is that they are able to aggregate smaller (asset) carriers' capacity to serve the shipper's needs. Often, shippers aims to maintain a manageable carrier base size (i.e., number of contracted carriers) across their networks. Rather than contracting with many, small carriers, they prefer to allow brokerages to manage these carriers. The benefit for smaller asset carriers of working with a broker or 3PL is that often they might not otherwise have access to some shippers' business. For larger asset carriers, the broker may provide opportunities to fill backhauls or moves to re-position a truck that does not have contracted volume and would otherwise be an empty, unpaid trip.

A shipper may set up a contract with a non-asset provider on a lane to lower the risk that loads go to the less predictable, volatile spot market. Brokers set prices with shippers for a contract term length and then typically pay (close to) spot market prices for their asset carriers over the course of the contract. They aim to hedge the market and, over time, still make a reasonably sustainable margin. When spot market prices are high, they may be paying carriers more than they are receiving from their contracted shippers. However when spot market prices are lower in a soft market, they receive higher payments from their fixed-price contracts than the price at which they are buying capacity on the market. 

Because of the difference between how asset and non-asset providers utilize markets, we predict non-asset carriers are more likely than asset carriers to be pulled from their contracted loads when spot prices are high and more attractive than contract prices. Thus they would be more responsive to index-based pricing.

Using an auction theory lens, this distinction in behaviors between how asset and non-asset carriers approach bid pricing is addressed in \cite{Scott2018}. We aim to expand on this by modeling the behaviors of asset and non-asset carriers separately.\footnote{Some carriers also offer both services. For example, large asset carriers J.B. Hunt, Schneider, and Knight-Swift all run a brokerage division of their business. In our dataset, we can distinguish between whether a company's asset or non-asset arm was awarded or tendered loads because they fall under different SCAC codes. SCACs, or Standard Carrier Alpha Codes, are unique 2- or 4-letter codes assigned by The National Motor Freight Traffic Association, Inc., (NMFTA) to identify transportation companies and are used throughout the freight industry for consistent carrier identification.} In this way, we extract each type of carriers' contract price stickiness for different freight and lane segments independently.

\subsection{Asset carrier fleet size} \label{sec:fleetsize}
Within the asset provider segment, carriers' cost structures vary by fleet size. For example, large carriers may be more able to absorb variations in market prices than would a smaller, owner-operator that owns a single truck and trailer. The distribution of asset carrier fleet size - both across the industry in the US and in our dataset - is highly skewed; about 60\% of total for-hire carriers in the US are independent owner-operators, and 96\% of fleets have fewer than 20 trucks. To account for this, we include the log of the carrier's fleet size (tractor count) as our measure of asset carrier size.

\begin{wrapfigure}[10]{l}{0.7\textwidth}
  \begin{center}
  \vspace{-\intextsep}
    \caption{Distribution of Asset Carrier Fleet Size}
    \label{fig:TractorCount}
    \includegraphics[width=0.7\textwidth]{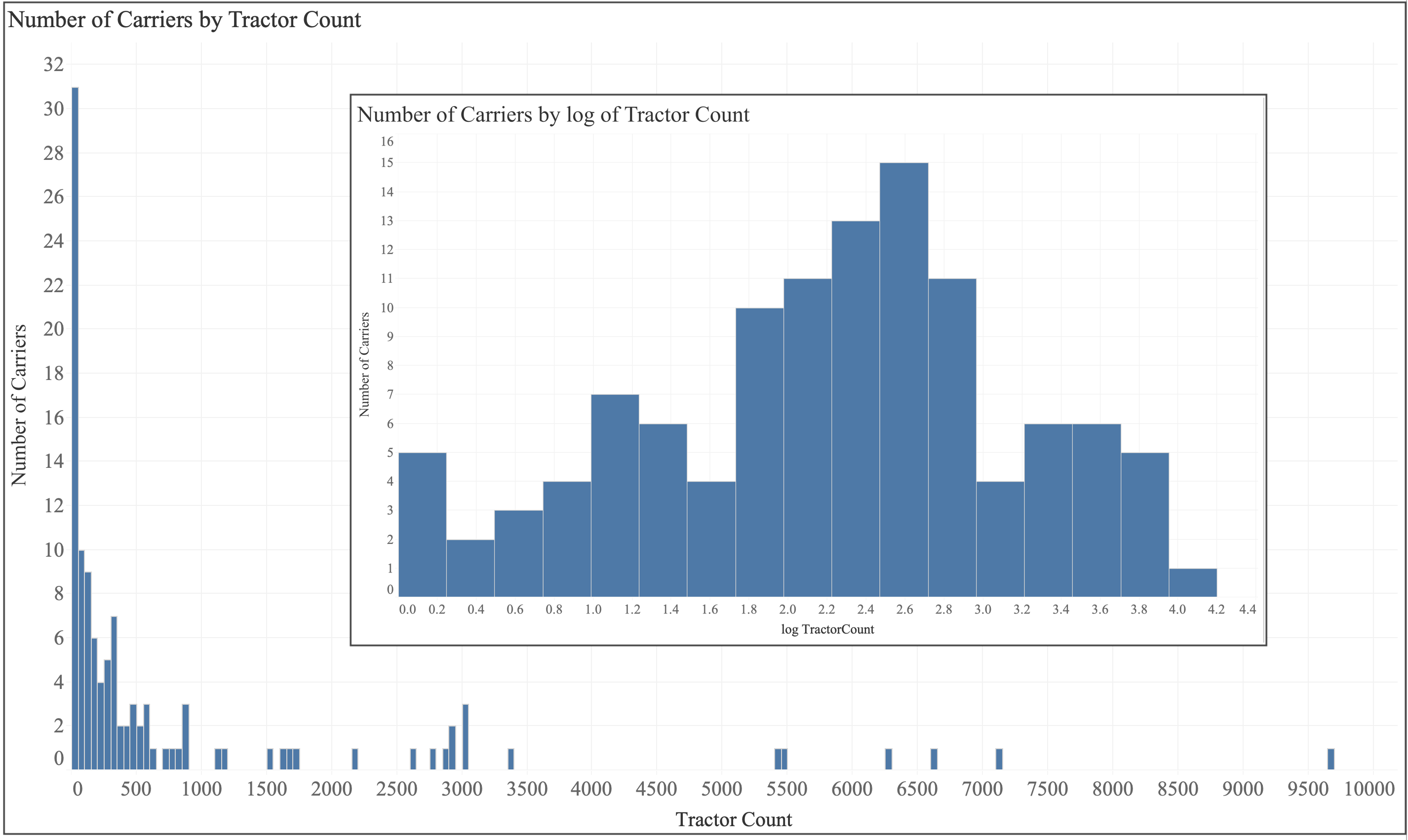}
  \end{center}
\end{wrapfigure} Figure \ref{fig:TractorCount} depicts the skewed distribution of carriers' number of tractors in the fleet and the normalized distribution of the log of carrier fleet size from our dataset.

\subsection{Other fixed effects} \label{sec:fixedeffects}
While we do not make formal hypotheses regarding the following variables, we do control for their fixed effects in our model and report their effects on carrier acceptance decisions. These include the shipper's size, measured by the log of its total monthly tendered volume across all lanes, and the shipper's industry vertical.
%which falls into one of the following five categories: (1) Automotive, (2) Food and Beverage (F\&B) and Consumer Packaged Goods (CPG), (3) Manufacturing, (4) Paper \& Packaging, and (5) Other.

Indicators for the lane's origin and destination regions are also included to account for their relative attractiveness. As discussed in Section \ref{sec:SRD} and \cite{Acocella2020a}, carrier acceptance is expected to vary across different inbound or outbound regions due to the business opportunities present at the origin and destination. We include in our carrier acceptance model a binary variable for each of the 15 regions of the US defined by general market demand patterns of our transportation management partner company, at both an origin and destination. %We choose the California region as our origin region indicator and the Ohio River region indicator as our destination to omit in the regression model, which are the regions with the highest outbound and inbound volume, respectively.

The results of our model indicate which load, lane, and carrier characteristics are statistically significant indicators for predicting primary carrier acceptance rate (PAR) at a load transaction level. Further, we address our carrier price stickiness research question by quantifying the changes in PAR as the contracted load price relative to the spot market price (i.e. SRD) changes for the resulting statistically significant freight and carrier segments.

\subsection{Market Condition}
Building off the research outlined in Section \ref{sec:litreview}, we expect carriers' contract price stickiness to differ depending on the general market condition. Our dataset spans two distinct market conditions: a soft market observed from the first week of February 2016 to the first week of July 2017 and again after the second week of January 2019 until the end of the time frame covered by the dataset; and a tight market observed before the first week of February 2016 and from the first week of July 2017 to the second week of January 2019. \cite{Acocella2020a} justify these market periods by identifying the weeks in which a statistically significant change is observed in the underlying structure of the TL freight market. We develop four distinct acceptance models: (1) asset carriers in soft markets, (2) non-asset carriers in soft markets, (3) asset carriers in tight markets, and (4) non-asset carriers in tight markets.

\section{Results} \label{sec:results}
In this section, we summarize our results by discussing the full GEE logistic regression models and use them to predict the (log) likelihood a load is accepted for each hypothesis. The regression results are tabulated in the electronic companion. The separate reported tables comprise a single model with the input variables described in Sections \ref{sec:SRD}-\ref{sec:fixedeffects}.

As validation of each of our four models, we report the Brier Score, which is the appropriate scoring metric when probability outcomes are the desired result \citep{Wallace2014, Niculescu2005, Zadrozny2002, Brier1950}. The Brier Score, $BS$, is the mean square error between the predicted probability an observation is in the the positive class, $p_n$ (here, an accepted load, $y_{c,k} =1$), and the actual outcome, $o_n \in \{0,1\}$:

\begin{equation} \label{eq:BrierScore}
    BS = \dfrac{1}{N}\sum^{N}_{n=1}(p_n-o_n)^2
\end{equation}

where $N$ is the total number of observations in the dataset. The Brier Score takes on values between 0 and 1. Better models have lower Brier Scores. 

To build our models, we segment each of the datasets for the four models into a training set on which we fit the model and a test set to measure model performance with a 70\%:30\% split. In each of the four pre-split datasets, the relative frequency of accepted loads is much higher than that of rejected loads. To develop unbiased models that do not naively favor accepted load predictions, we use a stratified sampling technique to define each of the training and test sets such that the ratio of the number of accepted loads to number of rejected loads in each split set is the same as that of the original pre-split dataset. In this way, the validated models can better predict carriers' load acceptance or rejection probabilities on new data.

\begin{wraptable}[7]{r}{5.5cm}
    \centering
    \vspace{-\intextsep}
    \caption{Test set Brier Scores}
    \label{tab:BrierScores}
    \begin{tabular}{c|c}
        Model & Brier Score \\ \hline \hline
        \shortstack{Asset carriers\\ Soft market} & 0.049 \\ \hline
        \shortstack{Asset carriers\\ Tight market} & 0.032 \\ \hline
        \shortstack{Non-asset carriers\\ Soft market} & 0.075 \\ \hline
        \shortstack{Non-asset carriers\\ Tight market} & 0.064 \\ \hline
    \end{tabular}
\end{wraptable} The Brier Scores for each of our four models on their respective test datasets are reported in Table \ref{tab:BrierScores}. Recall that the closer a score is to 0, the better. A ``good'' Brier Score largely depends on the dataset itself. However, we draw from the literature for reasonable comparison. \cite{Zadrozny2002} report test set Brier Scores for the best calibrated probability prediction models of five datasets ranging from 0.012 to 0.204. %0.095, 0.108 0.204, 0.0149, 0.0122
Similarly, \cite{Wallace2014} report calibrated model Brier Scores for 34 datasets between 0.042 to 0.319. All four of our models' Brier Scores sit below the average Brier Score value of the best fitting models in these previous studies. Thus, we conclude that our models perform well in predicting loads acceptance probability.

\subsection{Carrier price stickiness by service type and market condition} \label{sec:stickiness}
The results of the GEE logistic regression models are reported in Table \ref{tab:Sensitivity}. Hypothesis H1, which addresses carriers' overall contract price stickiness is supported: in general, primary carriers are more likely to accept a load with a contract price that is higher relative to the current, lane-specific spot market price. The statistically significant coefficients in Table \ref{tab:Sensitivity} show that as SRD increases and the spot price rises above contract prices, the probability a carrier accepts the load decreases for both carrier types in both market conditions. This suggests that in general, carriers can be incentivized away from their contracted loads as the external spot market increases relative to those contract prices.

\begin{figure*}[!hbt]
    \begin{center}
      \caption{Carrier Contract Price Stickiness Models}
      \label{fig:PriceSensitivity}
      \includegraphics[width=0.75\textwidth]{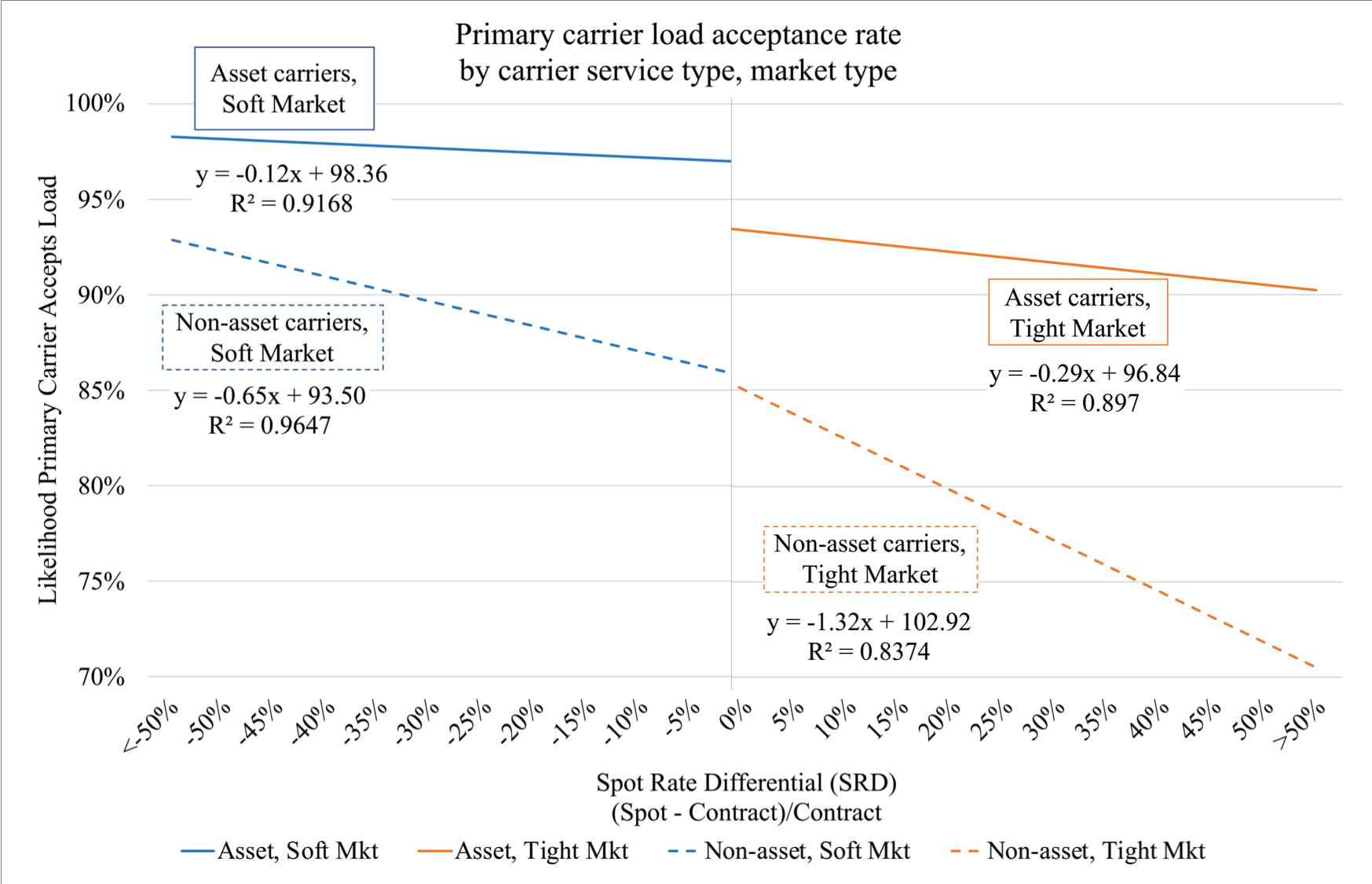}
    \end{center}
\end{figure*}

Figure \ref{fig:PriceSensitivity} illustrates this behavior for each carrier type and market condition. To construct the figure, we plot the probability a load is accepted at each of the statistically significant SRD values for a base case lane and fit a linear model to these plotted results./footnote{We show only the segments of these linear models over the SRD ranges that apply to the associated market condition. In other words, soft market models apply when SRD negative (spot market prices are below contract prices) and tight market models apply when SRD is positive.} The slope term of each linear model represents the carrier's contract price stickiness, or willingness to stick to contracted loads as spot market prices change relative to contract prices in the corresponding market condition.

Both carrier types are about twice as likely to be pulled from contract load prices in a tight market as they are in a soft market: asset carriers have a slope of -0.29 in tight markets as compared to -0.12 in a soft market and non-asset carriers have a slope of -0.65 in soft markets and -1.32 in tight markets. Moreover, H7 is supported: non-asset carriers are about five times more likely to be pulled from contracted loads than their asset-based counterparts in each market condition.

For shippers considering indexed contract pricing strategies, these results suggest that all else equal, non-asset carriers are more likely to respond to market-based pricing than asset carriers and that both carrier types may be more responsive to such pricing strategies in tight markets than soft.

Next, we discuss the results for specific freight and lane segments highlighted in our hypotheses. We demonstrate carriers' contract price stickiness by holding spot price constant at \$1000 for both soft and tight markets and plotting the contract price needed  to maintain a 90\% likelihood the primary carrier accepts the contracted load over the range of values each freight or lane segment of interest takes. We choose 90\% PAR because most shippers expect at least this level of service from their contracted carriers. Many may expect higher acceptance rates, however this threshold represents basic service level expectations in the shipper-carrier relationship. Our method of presenting contract price stickiness follows those discussed in Section \ref{sec:litreview} on general supplier price stickiness and responses to factors exogenous to the supply contract.\footnote{The figures identify segments where primary carriers require higher price incentive to stick to their contracts as the spot price approaches or surpasses the contract price. Contract prices of categorical variable values for which the GEE regression model coefficients are not statistically significant (see \ref{tab:Sensitivity}) are presented as the contract price associated with the baseline value.}

\subsection{Contract price stickiness: lane demand consistency} \label{sec:consistency}
Results for both lane tendering frequency and volatility are summarized in \ref{tab:Consistency}. They show that hypotheses H3 and H4 are supported particularly for asset carriers in both market conditions. Carriers prefer lanes with more frequently and consistently tendered volume; that is, carriers are more willing to stick to their contract prices on lanes with high frequency load tendering and lower week-to-week volume volatility.

Importantly, the shipper often cannot control its load tendering consistency; it is subject to external factors such as its own customers' demand patterns, inbound suppliers' schedules, and congestion on roadways or at ports. While the shipper does not control when its demand for trucks materializes, it does control what carriers are offered the loads when they do appear. Moreover, the shipper has historical knowledge of its lane demand patterns. Thus, the shipper's decision here is what bid prices to accept and how to tender loads that do appear for lanes with historically irregular cadence or inconsistent volumes.

\subsubsection{Tendering cadence}
The results weakly support Hypothesis H3: lanes on which loads are tendered less frequently see lower primary carrier acceptance. Moreover, both asset and non-asset carriers in both market conditions are less willing to stick to their contract prices for these low cadence lanes. (See \ref{tab:Consistency}.)

\begin{wrapfigure}[10]{r}{0.6\textwidth}
  \begin{center}
%   \vspace{-\intextsep}
    \caption{Contract Price Stickiness, Tender Cadence}
    \label{fig:Cadence}
    \includegraphics[width=0.58\textwidth]{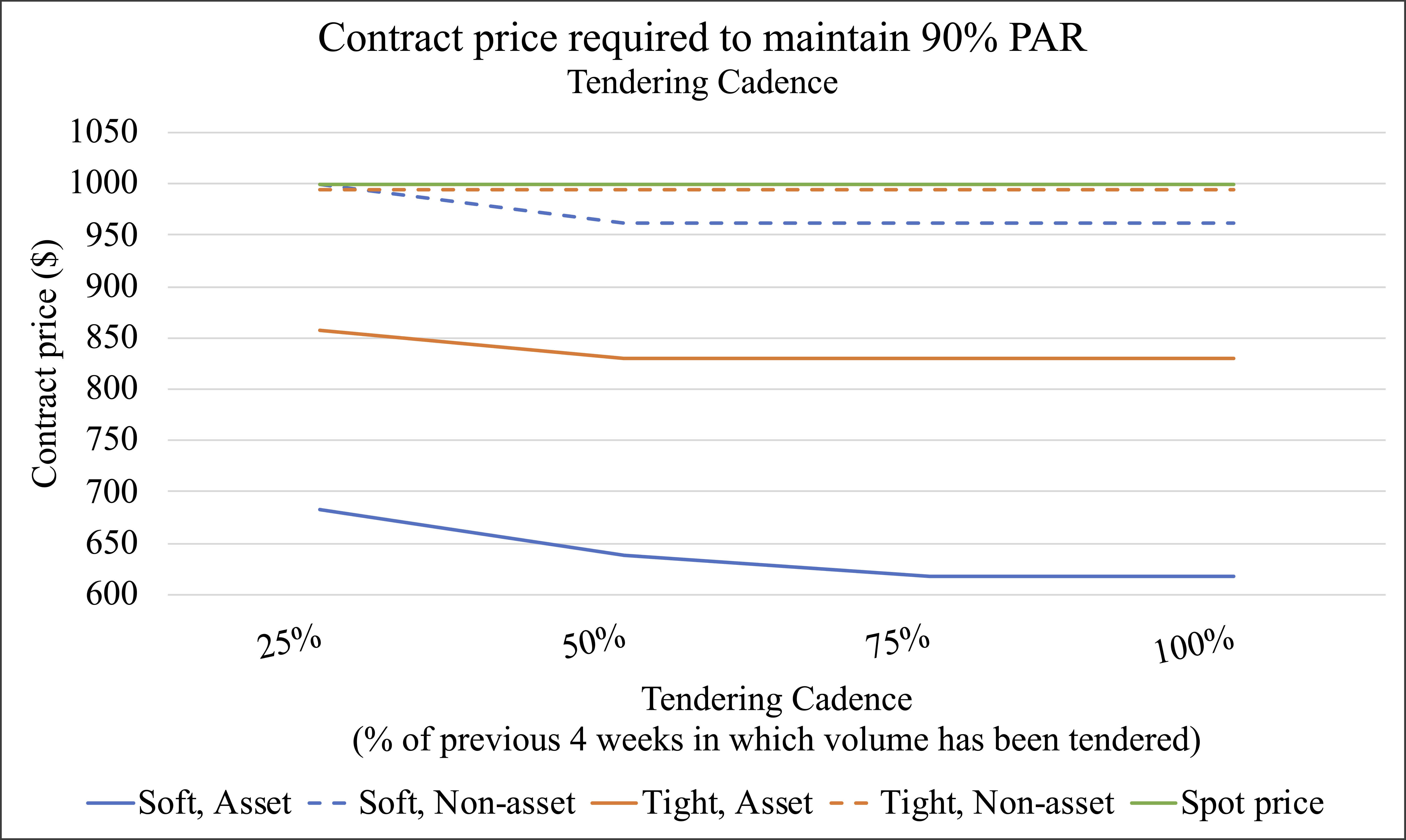}
  \end{center}
\end{wrapfigure}

Figure \ref{fig:Cadence} shows that as lane tendering cadence decreases, asset carriers' contract price needed to maintain 90\% PAR increases about 7\% in soft markets (to \$682 per load from \$617) and 2.5\% in tight markets (to \$857 from \$830).

Similarly, non-asset carriers require a 6\% contract price increase in soft markets to maintain 90\% PAR on low cadence lanes. While they do not appear to make load acceptance decisions based on lane tendering cadence in tight markets, these non-asset carriers set the highest contract prices for all cadence levels in tight markets. This may be because shippers often use non-asset brokerage services specifically for infrequently tendered lanes. These providers may be familiar with the undesirable, infrequently tendered lanes and knowingly set higher contract prices during the RFP. 

The results suggest that both carrier types are willing to stick with contract prices on moderate and high cadence lanes under both market conditions. However, lanes with very infrequent load tenders require higher contract prices relative to going spot market prices. Shippers may see better primary carrier acceptance on these low frequency lanes with market-based pricing strategies. Important to note, transportation practitioners and the literature both highlight that there must be enough business between the buyer and supplier (on a lane) to offset the additional effort involved in introducing either non-traditional, or more explicit contracts \citep{Scott2020, Sinha2019}.

\subsubsection{Tendered volume volatility}
\begin{wrapfigure}[9]{r}{0.65\textwidth}
  \begin{center}
  \vspace{-\intextsep}
    \caption{Contract Price Stickiness, Lane Volatility}
    \label{fig:Volatility}
    \includegraphics[width=0.57\textwidth]{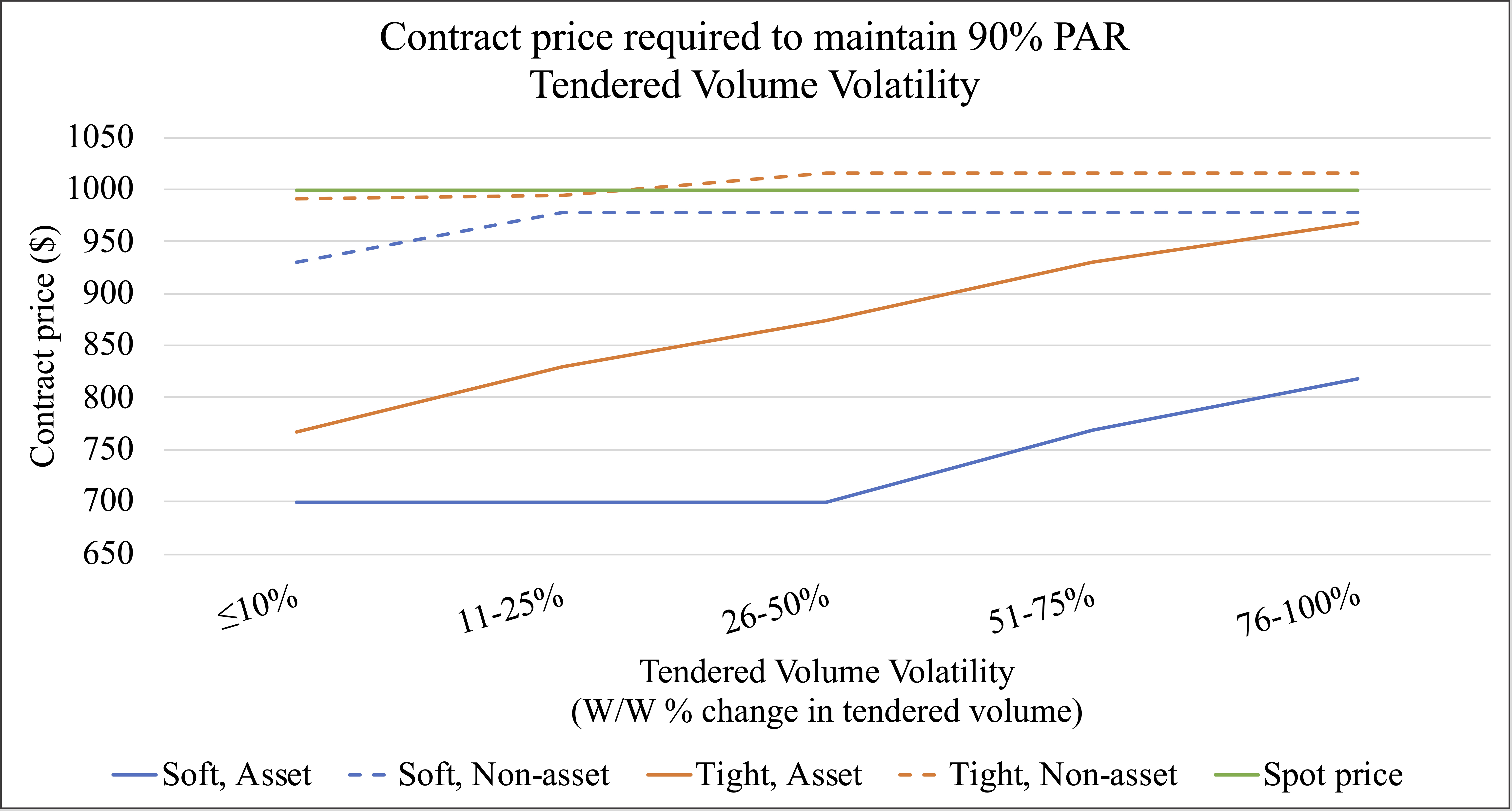}
  \end{center}
\end{wrapfigure}

Next, we consider the tendered volume volatility on a lane from a shipper to its primary carrier. Consistent with Hypothesis H4, higher tendered volatility leads to lower load acceptance probabilities, particularly for asset carriers. The GEE logistic regression model results are presented in Table \ref{tab:Consistency} and Figure \ref{fig:Volatility} below demonstrates carrier contract price stickiness. In soft markets, asset carriers stick to their contract prices for lanes with week-to-week tendered volume volatility up to 50\%.
\indent However, as tendering volatility increase, they are more likely to be pulled from the contract priced loads unless contract prices are much closer to the current spot market price. Specifically, contract prices needed to maintain 90\% acceptance are 17\% higher on these high volatility lanes than low volatility lanes in soft markets.

In tight markets, as compared to soft markets, even higher contract prices are needed for asset carriers to maintain high acceptance rates. Moreover, these price escalations begin at lower volatility lanes (i.e., 11-25\%). Contract prices needed on high volatility lanes are 26\% higher than those on low volatility lanes in tight markets. This suggests that in tight markets, asset carriers are more easily pulled from their contracts on lanes with lower tendering volatility than in soft markets.

Non-asset carriers are more willing to stick to their contract prices at high lane volatility in both market conditions as compared to their asset-based counterparts. Contract prices needed for non-asset providers to maintain 90\% load acceptance rates remain steady for lanes with more than 10\% week-over-week change in tendered volume, with an increase of only 5\% in soft markets for moderate and high volatility lanes.

Shippers should pay close attention to moderate and high volatility lanes in their networks. These lanes are particularly difficult for asset carriers to accommodate due to fixed capacities and a network of many customers' demand they continuously balance while non-asset providers do not. For the lanes on which a shipper's demand volatility is difficult to control or smooth out (by splitting the volume and tendering to multiple carriers, for example), it may be beneficial to introduce a market-based pricing strategy that adjusts contract load prices as spot market prices change to better incentivize primary asset carriers to accept loads.

\subsection{Contract price stickiness: surge volume} \label{sec:surge_sensitivity}
Table \ref{tab:Surge} summarizes the GEE logistic regression models' results demonstrating carriers' likelihood of accepting loads considered surge volume. Hypothesis H5 is strongly supported for asset carriers: asset primary carriers are less likely to accept loads that are above the lane awarded volume - particularly those over 20\% above awarded volume - than loads within their expected weekly tendered volume. Put another way, asset carriers are less willing to stick to their contract prices for excessive surge volume, especially in tight markets. \begin{wrapfigure}[10]{l}{0.6\textwidth}
  \begin{center}
  \vspace{-\intextsep}
    \caption{Contract Price Stickiness, Surge Volume}
    \label{fig:Surge}
    \includegraphics[width=0.58\textwidth]{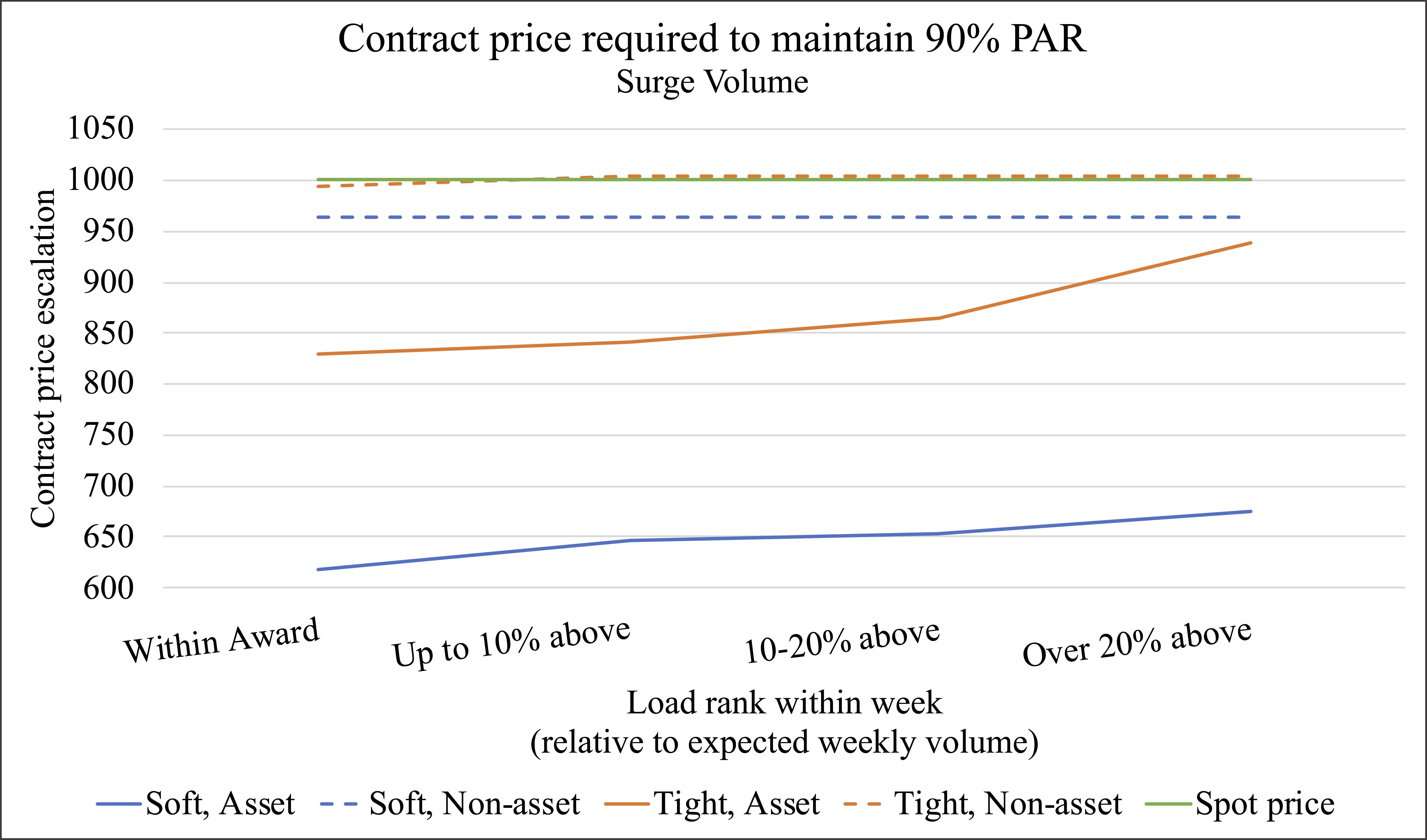}
  \end{center}
\end{wrapfigure} \indent Figure \ref{fig:Surge} shows that the contract price required to maintain 90\% acceptance for asset carriers in soft markets increases by 9\% for loads that are over 20\% above the awarded volume. These asset carrier contract prices further increase in tight markets. For surge volume over 20\% above the award in tight markets, shippers pay \$940 per load - almost at the \$1000 spot market price level - which equates to a 13\% increase in contract price from the awarded volume loads.

% \begin{figure}[!ht]
% 	\centering
% 	\begin{minipage}{.5\textwidth}
% 		\centering
% 	\caption{Contract Stickiness, Surge Volume}
%  \includegraphics[width=0.9\textwidth]{Surge_Volume.jpg}		\label{fig:Surge}
% 	\end{minipage}%
% 	\begin{minipage}{.5\textwidth}
% 		\centering
% 	\caption{Contract Stickiness, Lane Distance}
% \includegraphics[width=0.98624\textwidth]{Lane_Dist.jpg}		\label{fig:Distance}
% 	\end{minipage}
% \end{figure}

% \begin{figure}[!htb]
%     \begin{center}
%       \caption{Contract Price Stickiness, Surge Volume}
%       \label{fig:Surge}
%       \includegraphics[width=0.75\textwidth]{Surge_Volume.jpg}
%     \end{center}
% \end{figure}

The contract price required to maintain high acceptance rates for loads up to 10\% and 10-20\% above awarded volumes with asset carriers stays relatively steady. This moderate surge volume requires contract price increases of 4-6\% in both soft and tight markets. Asset carriers are willing to stick to their contract prices for these mid-level surge loads. This may be because shippers often communicate an expectation of primary carriers to accept anywhere from 10-20\% above the awarded volume without reducing acceptance or other carrier performance metrics in their (non-binding) contract service level agreements. Although asset carriers may have more difficulty providing this additional capacity than non-asset carriers, they may have factored some additional capacity into their strategic capacity allocation decisions during the original RFP. Surge loads over this 20\% threshold require higher contract prices, especially in tight markets, and asset carriers are more willing to be pulled from contracts for better priced alternatives on the spot market.

Non-asset carriers, on the other hand, do not appear to be incentivized away from contracts with higher priced alternatives on surge volume, even for loads that are more than 20\% above the awarded volume. This difference between asset and non-asset carriers is expected. Non-asset carriers are not limited in capacity in the way asset carriers are. While non-asset carriers must also balance supply and demand of capacity, they are better equipped to serve excess demand because they can access a large pool of asset carriers and aggregate capacity accordingly. Moreover, non-asset carriers already require higher contract prices to maintain 90\% acceptance for all volume - in fact, very close to the spot market price - regardless of surge classification. Thus, they may already be adequately incentivized to offer that capacity.

Shippers may see the most benefit from implementing indexed pricing applied to specific surge volume loads with asset carriers. This type of volume-based pricing strategy has been seen in practice. It is referred to as tier pricing, where levels of surge volume are set at a higher fixed price - determined during the strategic RFP - than that of the base awarded volume.

\subsection{Contract price stickiness: lane distance}
Table \ref{tab:Distance} summarizes the lane distance portion of the GEE logistic regression model results. It shows weak support of Hypothesis H2: non-asset carriers prefer longer lanes. In soft markets, loads on shorter lanes - specifically those with only a one-day travel time - are less likely to be accepted than all other lane distances. In tight markets, loads on longer lanes (three- and four-day lanes) are more likely to be accepted than loads on shorter lanes. However, there is no statistically significant difference in probability of load acceptance for asset carriers on different lane distances, suggesting that asset carriers do not consider lane distance in their load acceptance decision.

\subsection{Asset carrier contract price stickiness by fleet size} \label{sec:fleet}
The preceding sections consider carriers based on their service type: asset and non-asset. However, asset carriers range widely in their fleet sizes (Figure \ref{fig:TractorCount}), which directly impact their internal cost structures, tolerance to price fluctuations, access to network model optimization software to plan strategically, and as a result, willingness and ability to accept loads. Table \ref{tab:fleet} summarizes the GEE logistic regression models' results for an asset carriers' fleet size. Hypothesis H8 is not supported: all else equal, we do no find statistically significant evidence that the size of the asset carrier (measured here as log of Tractor Count) indicates its primary acceptance rate in soft or tight markets.

% \subsection{Fixed effects} \label{sec:resultsfixedeffects}
% Finally, we control for fixed effects in our model by including variables for origin and destination regions and shipper size and industry vertical. The results of the shipper fixed effects are in Table \ref{tab:shipperfixed} in Appendix \ref{AppendA}. Larger shippers typically experience higher freight acceptance rates from both asset and non-asset carriers in soft markets, all else equal. Moreover, a few of the shipper industry verticals observe statistically significant differences in PAR across market types. Of course a shipper's size and industry are not decisions or behaviors it can alter to incentivize carrier acceptance. However, the model results suggest that smaller shippers and those in specific industries may be more exposed to carrier rejections and may want to consider alternative relationships or pricing strategies with their carriers to improve observed load acceptance performance.

\section{Discussion and Implications for Index-based Pricing} \label{sec:implictions}
In this study we explore supplier price stickiness in response to customer demand patterns and exogenous market conditions. We take the case of firms that outsource their TL transportation service needs and aim to help them determine when, where, and with which transportation suppliers they should consider alternative pricing strategies from the standard, fixed-price contracts. In particular, we consider dynamic market-based pricing by measuring primary carriers' contract price stickiness, or the change in contract price relative to the going spot price needed so carriers maintain a reasonable load acceptance service level.

\textbf{1. Shippers should not delay or skip RFPs or mini-bids as markets tighten.} Both carrier types (asset and non-asset) are twice as likely to stick to their contract priced loads in a soft market than a tight market. Shippers can use this to inform the timing of their strategic procurement event as market conditions may be changing. For example, say a shipper runs its annual strategic RFP in January. In the preparation months leading up to the event, the shipper may observe that spot market prices or other leading indicators of general market conditions are rising. Some shippers in this situation may consider delaying the RFP and keeping the current contracted prices that had been set during a softer, lower-priced market rather than opening itself up for carriers to lock in higher prices. Alternatively, the shipper may choose to run the RFP anyway, perhaps allowing for slightly increased contract prices, and assume it can expect good primary carrier acceptance rates for the next year even as markets further tighten. Finally, the shipper may take a middle-ground approach and proactively increase rates for certain core carriers on specific important lanes to keep them out of an RFP.

However, the results of our study suggest that no matter the decision, if spot prices continue to rise as the market tightens further, the contract prices will become less competitive (i.e., their spot rate differential will increase) and primary carriers with low contract price stickiness will begin opportunistically rejecting loads at higher rates, defecting to either fresher contract prices with other shippers or to the spot market. Therefore, when markets are tightening, shippers need to continuously ensure their contract prices stay competitive with the going spot market prices, regardless of their RFP timing. One approach shippers can take as market conditions become more constrained is to execute smaller, focused, more frequent ``mini-bids'' with core primary carriers on the most important, susceptible lanes. These mini-bids result in shorter term (e.g., 30-, 60-, or 90-day) contracts. Another approach is to implement a market-based pricing strategy, as we propose in this study, which effectively ensures contract prices are up-to-date.

\textbf{2. Non-asset carriers are best suited for market-based contracts.} All else equal, asset primary carriers are five times more likely to stick to their contract priced loads than non-asset primary carriers, or brokers, in both soft and tight markets. In fact, across most network and freight segments, shippers must pay these non-asset primary carriers just about spot market prices to maintain high load acceptance rates. This suggests that brokers may respond to index-based contracts better than asset-based providers.

\textbf{3. Market-based pricing shows promise for volatile demand, low frequency lanes, and surge volume.} Asset primary carriers are less likely to stick to their contract priced loads on lanes with infrequently tendered loads and high week-to-week volatility of tendered volume. Shippers often have little control over when loads actually materialize and carriers' capacity is needed. However, they do know historical demand patterns on their lanes. They can control which carriers they tender the loads that do materialize and the pricing strategies they are willing to implement. Our results suggest a market-based pricing strategy for these low cadence or high volatility lanes would better incentivize asset carriers to stick to their contracts.
% However, for low cadence lanes, there must be enough materialized volume over the course of the contract for the intended benefits of an alternative contract to be felt.

In addition to these lanes types, shippers can expect asset carriers to respond to market-based pricing for surge volume - in other words, loads that are in excess of the expected weekly volume communicated during the RFP. This finding relates to the two stages of TL transportation. During the first stage, the strategic RFP, the shipper communicates the expected weekly volume on each lane that the carriers bid on. Once the carrier wins the lane, it plans to allocate capacity for that expected volume, perhaps with an additional 10-20\% buffer. It also uses this knowledge of expected demand to balance the demand for trucks across its existing network and to inform its bid strategy in other shippers' RFPs. However, during the second stage, the operational stage when loads materialize and are tendered to that carrier, the actual demand may well exceed the capacity for which the carrier has planned. Asset carriers in particular are constrained in their total capacity available at any one time. To maintain high acceptance rates of loads that exceed 20\% over the awarded volume on a lane, carriers require a price incentive. This is an important segment of shippers' freight: consistently across our time horizon, 10-15\% of total tendered volume is in this surge category.

In fact, a tier pricing strategy where surge volume is priced higher than the baseline awarded volume is used by some shippers and their larger or core carriers, precisely for the reasons described. However, these tier-based rates are still fixed at the time the contract is established. A dynamic market-based price, on the other hand, would ensure the tiered price incentive remains competitive with the current market.

\textbf{4. Indexed pricing may result in lower load acceptance if applied in soft markets.} Important to note that we discuss market-based pricing with the implication that contracted load prices increase as market prices increase in a tight market, but also decrease as prices decrease in soft markets - in other words, ``symmetric'' indexing. Shippers may want to consider using dynamic market pricing as an incentive for primary carriers whereby indexed contract prices only increase as the market tightens, but settles to the competitive, fixed contract price when the market prices are decreasing. Otherwise, for segments that require a contract price \textit{premium} (i.e., higher contract price than spot market price) for high primary carrier acceptance, the shipper would expect to see decreased primary carrier acceptance when the indexed price symmetrically decreases.

Our study suggests that there is still a place for the widely used fixed-price contracts. On lanes where tendered volume is consistent and frequent, both shippers and carriers prefer a set price to plan and budget around. While these prices may also become outdated as spot market prices increase, carriers are more willing to stick to their contracts for attractive freight.\footnote{While low tight market period primary carrier acceptance rates are undesirable, the majority of freight still moves under fixed contract rates even in tight markets. However, we find evidence that a market-based approach would improve primary carrier acceptance rates.} Moreover, competitive fixed price contracts may still be best suited for soft market conditions.

\section{Limitations and Future Research} \label{sec:futureresearch}
Our empirical modeling results offer both academic and practical contributions. First, we add to the econometric literature on supplier price stickiness in response to demand and exogenous market dynamics. We do so by quantifying TL transportation suppliers' price stickiness as market prices change and for different customer demand patterns, freight segments, characteristics of shippers' networks, and supplier service types. Our work explores real-time decision that previous literature overlooks; transportation suppliers have the spot market as a dynamic alternative option to their contracted business for every transaction.

In addition, we add to the literature specific to shipper and carrier relationships, particularly with a focus on how changing market conditions impact behaviors. We do so by utilizing a uniquely extensive and detailed dataset. Previous empirical studies in the space have been limited in that their data represent only a single shipper's business, and often contain limited or no tendering or carrier acceptance service level information.

This study is not without its limitations. Precisely due to the observational nature of our dataset, we cannot control for outside influences on carriers' acceptance decisions. We attempt to account for these factors by segmenting the contract price stickiness analysis by lane, load, and carrier types and other fixed effects that have been previously identified as contributing factors. In doing so, we build models with good performance (i.e., very low Brier Scores). However, there may still be alternative explanations for carriers' acceptance decisions.

Notwithstanding these limitations, our study further adds to the existing shipper-carrier relationship literature by isolating the impact of market prices on the way carriers make freight acceptance decision for their contracted shippers. As discussed in the motivation of this research, there has been growing interest in index-based pricing in the TL industry. This research serves as a starting point. We demonstrate where shippers can expect to see carrier acceptance behaviors most influenced by dynamic, market-based pricing methods. An interesting future stream of literature could develop strategies for shippers and carriers to design and implement these freight contracts.

\ACKNOWLEDGMENT{The authors would like to thank the anonymous reviewers for greatly improving the quality of this manuscript.}

%%REFERENCES%%
%%%%%%%%%%%%%%%%%%%%%%%%%%%%%%%%%%%%%%%%%%%%%%%%%%%%%%%%%%%%%%%%%%%%%%%%%%%%%%%%%%%%%%%%%%%%%%%%%%%%%%%%%%%%%%%%%%%%%%%%%%%%%%%%%%%%
%% This template complies references using bibtex. You will need to use pomsref.bst file for biblography style.
%REFERENCES USING BIBTEX FILES
%%%%%%%%%%%%%%%%%%%%%%%%%%%%%%%%%%%%%%%%%%%%%%%%%%%%%%%%%%%%%%%%%%%%%%%%%%%%%%%%%%%%%%%%%%%%%%%%%%%%%%%%%%%%%%%%%%%%%%%%%%%%%%%%%%%%

\bibliographystyle{pomsref} 

 \let\oldbibliography\thebibliography
 \renewcommand{\thebibliography}[1]{%
 	\oldbibliography{#1}%
 	\baselineskip8pt %Change this for line spacing within the same reference
 	\setlength{\itemsep}{8pt}% %Change this for spacing between two referneces
 }
\bibliography{ref2} 

\begin{thebibliography}{63}
\expandafter\ifx\csname natexlab\endcsname\relax\def\natexlab#1{#1}\fi
\expandafter\ifx\csname url\endcsname\relax
  \def\url#1{{\tt #1}}\fi
\expandafter\ifx\csname urlprefix\endcsname\relax\def\urlprefix{URL }\fi
\expandafter\ifx\csname urlstyle\endcsname\relax
  \expandafter\ifx\csname doi\endcsname\relax
  \def\doi#1{doi:\discretionary{}{}{}#1}\fi \else
  \expandafter\ifx\csname doi\endcsname\relax
  \def\doi{doi:\discretionary{}{}{}\begingroup \urlstyle{rm}\Url}\fi \fi

\bibitem[{Acocella et~al.(2020)Acocella, Caplice, and Sheffi}]{Acocella2020a}
Acocella, Angela, Chris Caplice, Yossi Sheffi. 2020.
\newblock Elephants or goldfish?: An empirical analysis of carrier reciprocity
  in dynamic freight markets.
\newblock {\it Transportation Research Part E: Logistics and Transportation
  Review\/}, { 142} 102073.

\bibitem[{Aemireddy and Yuan(2019)}]{Yuan2019}
Aemireddy, Nishitha~Reddy, Xiyang Yuan. 2019.
\newblock Root cause analysis and impact of unplanned procurement on truckload
  transportation costs.
\newblock M. Eng in Logistics Thesis, MIT.

\bibitem[{Amiryan and Bhattacharjee(2015)}]{Amiryan2015}
Amiryan, Nane, Sharmistha Bhattacharjee. 2015.
\newblock Relationship between price and performance: An analysis of the us
  trucking market.
\newblock M. Eng in Logistics Thesis, MIT.

\bibitem[{Apel et~al.(2005)Apel, Friberg, and Hallsten}]{Apel2005}
Apel, Mikael, Richard Friberg, Kerstin Hallsten. 2005.
\newblock Microfoundations of macroeconomic price adjustment: Survey evidence
  from swedish firms.
\newblock {\it Journal of Money\/}, { 37} 313–338.

\bibitem[{Ballinger(2004)}]{Ballinger2004}
Ballinger, Gary~A. 2004.
\newblock Using generalized estimating equations for longitudinal data
  analysis.
\newblock {\it Organizational Research Methods\/}, { 7} (2), 127-150.

\bibitem[{Ballou(1991)}]{Ballou1991}
Ballou, Ronald. 1991.
\newblock The accuracy in estimating truck class rates for logistical planning.
\newblock {\it Transportation Research - Part A: Policy and Practice\/}, { 25}
  (6), 327-337.

\bibitem[{Barnes-Schuster et~al.(2002)Barnes-Schuster, Bassok, and
  Anupindi}]{Barnes2002}
Barnes-Schuster, Dawn, Yehuda Bassok, Ravi Anupindi. 2002.
\newblock Coordination and flexibility in supply contracts with options.
\newblock {\it Manufacturing \& Service Operations Management\/}, { 4} (3),
  171-207.

\bibitem[{Belzer and Sedo(2018)}]{Belzer2018}
Belzer, Michael~H., Stanley~A. Sedo. 2018.
\newblock Why do long distance truck drivers work extremely long hours?
\newblock {\it Economic and Labour Relations Review\/}, { 29} (1), 59-79.

\bibitem[{Blinder(1991)}]{Blinder1991}
Blinder, Alan~S. 1991.
\newblock Why are prices sticky? preliminary results from an interview study.
\newblock {\it American Economic Review\/}, { 81} 89-96.

\bibitem[{Blinder et~al.(1998)Blinder, Canetti, Lebow, and Rudd}]{Blinder1998}
Blinder, Alan~S., Elie Canetti, David Lebow, Jeremy Rudd. 1998.
\newblock Asking about prices: A new approach to understanding price
  stickiness.
\newblock {\it The Journal of Consumer Affairs\/}, { 32} (2), 424-427.

\bibitem[{{BR Williams}(2020)}]{BRWilliams2020}
{BR Williams}. 2020.
\newblock What is ``index-based pricing'' in transportation and logistics and
  how can it help my organization?
\newblock
  \urlprefix\url{https://www.brwilliams.com/blog/what-is-index-based-pricing-in-transportation-and-logistics/}.
\newblock Accessed March 29, 2021.

\bibitem[{Brier(1950)}]{Brier1950}
Brier, Glenn~W. 1950.
\newblock Verification of forecasts expressed in terms of probability.
\newblock {\it U.S. Department of Commerce Monthly Weather Review\/}, { 78}
  1-3.

\bibitem[{Burks and Monaco(2019)}]{Burks2019}
Burks, Stephen, Kristen Monaco. 2019.
\newblock Is the u.s. labor market for truck drivers broken.
\newblock {\it Monthly Labor Review\/}, { 142} (3), 1-23.

\bibitem[{Cachon(2003)}]{Cachon2003}
Cachon, G{'e}rard~P. 2003.
\newblock Supply chain management: Design, coordination, and operation.
\newblock {\it Handbooks in Operations Research and Management Science\/}, {
  11} 227-339.

\bibitem[{Cachon and Lariviere(2005)}]{Cachon2005}
Cachon, G{'e}rard~P., Martin~A. Lariviere. 2005.
\newblock Supply chain coordination with revenue-sharing contracts: Strengths
  and limitations.
\newblock {\it Management Science\/}, { 51} (1), 30-44.

\bibitem[{Cao et~al.(2012)Cao, Dong, and Tomlin}]{Cao2012}
Cao, Shutao, Wei Dong, Ben Tomlin. 2012.
\newblock The sensitivity of producer prices to exchange rates : Insights from
  micro data.
\newblock Tech. Rep. 2012-20, Bank of Canada Working Paper.

\bibitem[{Caplice(2007)}]{Caplice2007}
Caplice, Chris. 2007.
\newblock Electronic markets for truckload transportation.
\newblock {\it Production and Operations Management\/}, { 16} (4), 423-436.

\bibitem[{Cecchetti(1986)}]{Cecchetti1986}
Cecchetti, Stephen~G. 1986.
\newblock The frequency of price adjustment: A study of the newsstand prices of
  magazines.
\newblock {\it Journal of Econometrics\/}, { 31} 255–274.

\bibitem[{{C.H. Robinson}(2015)}]{CHR2015}
{C.H. Robinson}. 2015.
\newblock Do `favored shippers' really receive better pricing and service?
\newblock White Paper.

\bibitem[{{Convoy}(2020)}]{Convoy2020}
{Convoy}. 2020.
\newblock A new approach to primary freight.
\newblock \urlprefix\url{https://convoy.com/white-paper-guaranteed-primary/}.
\newblock Industry White Paper, Accessed April 1, 2021.

\bibitem[{Daganzo(2005)}]{Daganzo2004}
Daganzo, Carlos. 2005.
\newblock Transportation costs. chap. 2.3.
\newblock {\it Logistics Systems Analysis\/}. 23-29.
\newblock 4th ed.

\bibitem[{Deneckere et~al.(1997)Deneckere, Marvel, and Peck}]{Deneckere1997}
Deneckere, Raymond, Howard~P. Marvel, James Peck. 1997.
\newblock Demand uncertainty and price maintenance: Markdowns as destructive
  competition.
\newblock {\it The American Economic Review\/}, { 87} (4), 619-641.

\bibitem[{Fabiani et~al.(2006)Fabiani, Druant, Hernando, Kwapil, Landau,
  Loupias, Martins, Matha, Sabbatini, Stahl, and Stokman}]{Fabiani2006}
Fabiani, Silvia, Martine Druant, Ignacio Hernando, Claudia Kwapil, Bettina
  Landau, Claire Loupias, Fernando Martins, Thomas Matha, Roberto Sabbatini,
  Harald Stahl, Ad~C.~J. Stokman. 2006.
\newblock What firms’ surveys tell us about price-setting behavior in the
  euro area.
\newblock {\it International Journal of Central Banking\/}, { 2} 3-48.

\bibitem[{Fabiani et~al.(2007)Fabiani, Loupias, Martins, Sabbatini
  et~al.}]{Fabiani2007}
Fabiani, Silvia, Claire~Suzanne Loupias, Fernando Manuel~Monteiro Martins,
  Roberto Sabbatini, et~al. 2007.
\newblock {\it Pricing decisions in the euro area: how firms set prices and
  why\/}.
\newblock OUP USA.

\bibitem[{Greene(2003)}]{Greene2003}
Greene, William~H. 2003.
\newblock {\it Econometric Analysis\/}.
\newblock 5th ed. Pearson Education.

\bibitem[{Hall et~al.(2000)Hall, Walsh, and Yates}]{Hall2000}
Hall, Simon, Mark Walsh, Anthony Yates. 2000.
\newblock Are uk companies’ prices sticky?
\newblock {\it Oxford Economic Papers\/}, { 52} 425–446.

\bibitem[{Hardin and Hilbe(2012)}]{Hardin2012}
Hardin, James, Joseph Hilbe. 2012.
\newblock {\it Generalized Estimating Equations (GEE)\/}, vol.~99.
\newblock Chapman \& Hall.
\newblock \doi{10.1002/0470013192.bsa250}.

\bibitem[{Harding(2005)}]{Harding2005}
Harding, Matthew. 2005.
\newblock Can shippers and carriers benefit from more robust transportation
  planning methodologies?
\newblock M. Eng in Logistics Thesis, MIT.

\bibitem[{Heide and Miner(1992)}]{Heide1992}
Heide, Jan~B., Anne~S. Miner. 1992.
\newblock The shadow of the future: Effects of anticipated interaction and
  frequency of contact on buyer-seller cooperation.
\newblock {\it Academy of Management Journal\/}, { 35} (2), 265-291.

\bibitem[{{J.B. Hunt}(2015)}]{JBHunt}
{J.B. Hunt}. 2015.
\newblock 660 minutes: How improving driver efficiency increases capacity.
\newblock Industry White Paper.

\bibitem[{Kim(2013)}]{Kim2013}
Kim, Yoo~Joon. 2013.
\newblock Analysis of truckload prices and rejection rates.
\newblock M. Eng in Logistics Thesis, MIT.

\bibitem[{Lariviere(1999)}]{Lariviere1999}
Lariviere, Martin~A. 1999.
\newblock Supply chain contracting and coordination with stochastic demand.
\newblock Sridhar Tayur, Ram Ganeshan, Michael Magazine, eds., {\it
  Quantitative Models for Supply Chain Management\/}, chap.~8. Springer,
  233-268.

\bibitem[{Liang and Zeger(1986)}]{Liang1986}
Liang, Kung-Yee, Scott~L. Zeger. 1986.
\newblock Longitudinal data analysis using generalized linear models.
\newblock {\it Biometrika\/}, { 73} (1), 13-22.

\bibitem[{Lindsey and Mahmassani(2015)}]{Lindsey2015b}
Lindsey, C, H~S Mahmassani. 2015.
\newblock Measuring carrier reservation prices for truckload capacity in the
  transportation spot market choice experiment.
\newblock {\it Transportation Research Record\/}, { 2478} 123-132.

\bibitem[{Loupias and Ricart(2006)}]{Loupias2006}
Loupias, Claire, Roland Ricart. 2006.
\newblock Price setting in the french manufacturing sector: New evidence from
  survey data.
\newblock {\it Revue d’Economie Politique\/}, { 116} 541–554.

\bibitem[{Loupias and Sevestre(2013)}]{Loupias2013}
Loupias, Claire, Patrick Sevestre. 2013.
\newblock Costs, demand, and producer price changes.
\newblock {\it Review of Economics and Statistics\/}, { 95} (1), 315-327.

\bibitem[{Masten(2010)}]{Masten2009}
Masten, Scott~E. 2010.
\newblock Long-term contracts and short-term commitment: Price determination
  for heterogeneous freight transactions.
\newblock {\it American Law and Economics Review\/}, { 11} (1), 79-111.

\bibitem[{Morris and Joyce(1988)}]{Morris1988}
Morris, Michael~H., Mary~L. Joyce. 1988.
\newblock How marketers evaluate price sensitivity.
\newblock {\it Industrial Marketing Management\/}, { 17} (2), 169-176.

\bibitem[{Niculescu-Mizil and Caruana(2005)}]{Niculescu2005}
Niculescu-Mizil, Alexandru, Rich Caruana. 2005.
\newblock Predicting good probabilities with supervised learning.
\newblock {\it Proceedings of the 22nd international conference on Machine
  learning\/}. 625-632.

\bibitem[{Oxley(2012)}]{Lumineau2012}
Oxley, Fabrice Lumineau Joanne~E. 2012.
\newblock Let's work it out (or we'll see you in court): Litigation and private
  dispute resolution in vertical exchange relationships.
\newblock {\it Organization Science\/}, { 23} (3), 820-834.

\bibitem[{Pasternack(1985)}]{Pasternack1985}
Pasternack, Barry~Alan. 1985.
\newblock Optimal pricing and return policies for perishable commodities.
\newblock {\it Marketing Science\/}, { 4} (2), 166-176.

\bibitem[{Pickett(2018)}]{Coyote2018}
Pickett, Chris. 2018.
\newblock The coyote curve: A model for mitigating risk and uncertainty in
  modern supply chain operations.
\newblock Coyote Logistics.

\bibitem[{Rinehart et~al.(2004)Rinehart, Eckert, Handfield, Page, and
  Atkin}]{Rinehart2004}
Rinehart, Lloyd~M, James~A Eckert, Robert~B Handfield, Thomas~J Page, Thomas
  Atkin. 2004.
\newblock An assessment of supplier-customer relationships.
\newblock {\it Journal of Business Logistics\/}, { 25} (1), 25-62.

\bibitem[{{Schneider Transportation}(2019)}]{Schneider2019}
{Schneider Transportation}. 2019.
\newblock Management optimizes packaging manufacturer's spot freight with index
  pricing model.
\newblock
  \urlprefix\url{https://schneider.com/resources/case-study/how-market-index-pricing-model-optimizes-spot-freight-shipping}.
\newblock Industry White Paper.

\bibitem[{Scott(2015)}]{Scott2015}
Scott, Alex. 2015.
\newblock The value of information sharing for truckload shippers.
\newblock {\it Transportation Research Part E: Logistics and Transportation
  Review\/}, { 81} 203-214.

\bibitem[{Scott(2018)}]{Scott2018}
Scott, Alex. 2018.
\newblock Carrier bidding behavior in truckload spot auctions.
\newblock {\it Journal of Business Logistics\/}, { 39} (4), 267–281.

\bibitem[{Scott et~al.(2020)Scott, Craighead, and Parker}]{Scott2020}
Scott, Alex, Christopher Craighead, Chris Parker. 2020.
\newblock Now you see it, now you don't: Explicit contract benefits in
  extralegal exchanges.
\newblock {\it Production and Operations Management\/}, { 29} (6), 1467-1486.

\bibitem[{Scott et~al.(2016)Scott, Parker, and Craighead}]{Scott2016}
Scott, Alex, Chris Parker, Christopher Craighead. 2016.
\newblock Service refusals in supply chains: Drivers and deterrents of freight
  rejection.
\newblock {\it Transportation Science\/}, { 51} (4), 1086-1101.

\bibitem[{Simchi-Levi et~al.(2014)Simchi-Levi, Chen, and
  Bramel}]{SimchiLevi2014}
Simchi-Levi, David, Xin Chen, Julien Bramel. 2014.
\newblock {\it The Logic of Logistics\/}.
\newblock 3rd ed. Springer.

\bibitem[{Singh(2021)}]{Singh2021}
Singh, Omar. 2021.
\newblock Perfectly executed routing guides fail 100\% of the time.
\newblock
  \urlprefix\url{https://www.freightwaves.com/news/3-reasons-perfectly-executed-routing-guides-fail-100-of-the-time}.
\newblock Surge Transportation.

\bibitem[{Singh et~al.(2016)Singh, Billa, and Arora}]{Deloitte2016}
Singh, Ranjit, Ramesh Billa, Bhupinder~Singh Arora. 2016.
\newblock Index based pricing: Managing risk and profitability.
\newblock {\it Deloitte\/}, .

\bibitem[{Sinha and Thykandi(2019)}]{Sinha2019}
Sinha, Atmaja, Rakesh Thykandi. 2019.
\newblock Alternate pricing model for transportation contracts.
\newblock M. Eng in Logistics Thesis, MIT.

\bibitem[{Sokoloff and Zhang(2020)}]{Sokoloff2020}
Sokoloff, David, Gaohui Zhang. 2020.
\newblock Predicting and planning for the future: North american truckload
  transportation.
\newblock M. Eng in Logistics Thesis, MIT.

\bibitem[{Telser(1980)}]{Telser1980}
Telser, Lester~G. 1980.
\newblock A theory of self-enforcing agreements.
\newblock {\it The Journal of Business\/}, { 53} (1), 27-44.

\bibitem[{{Uber Freight}(2021)}]{Uber2021}
{Uber Freight}. 2021.
\newblock Uber freight expands procurement solutions with market access.
\newblock \urlprefix\url{http://www.uber.com/blog/uber-freight-market-access/}.
\newblock Accessed March 29, 2021.

\bibitem[{Wallace and Dahabreh(2014)}]{Wallace2014}
Wallace, Bryon~C., Issa~J. Dahabreh. 2014.
\newblock Improving class probability estimates for imbalanced data.
\newblock {\it Knowledge and Information Systems\/}, { 41} 33-52.

\bibitem[{Wathne and Heide(2000)}]{Wathne2000}
Wathne, Kenneth~H., Jan~B. Heide. 2000.
\newblock Opportunism in interfirm relationships: Forms, outcomes, and
  solutions.
\newblock {\it Journal of Marketing\/}, { 64} (4), 36-51.

\bibitem[{Williamson(1975)}]{Williamson1975}
Williamson, Oliver. 1975.
\newblock {\it Markets and Hierarchies: Analysis and Antitrust Implications\/}.
\newblock Free Press, New York.

\bibitem[{Williamson(1985)}]{Williamson1985}
Williamson, Oliver. 1985.
\newblock {\it The Economic Institutions of Capitalism: Firms, Markets,
  Relational Contracting\/}.
\newblock Free Press, New York.

\bibitem[{Williamson(1979)}]{Williamson1979}
Williamson, Oliver~E. 1979.
\newblock Transaction-cost economics: The governance of contractual relations.
\newblock {\it The Journal of Law and Economics\/}, { 22} (2), 233-262.

\bibitem[{Williamson(2002)}]{Williamson2002}
Williamson, Oliver~E. 2002.
\newblock The theory of the firm as governance structure: From choice to
  contract.
\newblock {\it The Journal of Economic Perspectives\/}, { 16} (3), 171-195.

\bibitem[{Zadrozny and Elkan(2002)}]{Zadrozny2002}
Zadrozny, Bianca, Chrles Elkan. 2002.
\newblock Transforming classifier scores into accurate multiclass probability
  estimates.
\newblock {\it Proceedings of the ACM SIGKDD International Conference on
  Knowledge Discovery and Data Mining\/}. 694-699.

\bibitem[{Zsidisin et~al.(2007)Zsidisin, Voss, and Schlosser}]{Zsidisin2007}
Zsidisin, George, Douglas Voss, Matt Schlosser. 2007.
\newblock Shipper-carrier relationships and their effect on carrier
  performance.
\newblock {\it Transportation Journal\/}, { 46} (2), 5-18.

\end{thebibliography}

%% Here starts the e-companion (EC). Place your appendix content here. 
%%%%%%%%%%%%%%%%%%%%%%%%%%%%%%%%%%%%%%%%%%%%%%%%%%%%%%%%%%
\ECSwitch % Comment this line out if you do not need e-companion.
%%%%%%%%%%%%%%%%%%%%%%%%%%%%%%%%%%%%%%%%%%%%%%%%%%%%%%%%%%

%%% Main head for the e-companion
\ECHead{E-Companion for ``The end of `set it and forget it' pricing? Opportunities for market-based freight contracts''}

\noindent This e-companion reports the detailed results of the logistic regression model, which is used to quantify carriers' contract price stickiness for segments indicated by model variables.

% \begin{appendices}
\begin{table}[!h]
    \centering
    \caption{GEE model results: Spot Rate Differential, asset and non-asset carriers, soft and tight markets}
    \label{tab:Sensitivity}
    \small
    \begin{tabular}{l|c|c|c|c}
        \shortstack{Spot Rate\\ Differential} & \shortstack{Asset carriers\\ Soft market} & \shortstack{Asset carriers\\ Tight market} & \shortstack{Non-asset carriers\\ Soft market} & \shortstack{Non-asset carriers\\ Tight market}\\\hline\hline
        Constant & 2.0032 & 1.2945 & 0.634 & 1.0894 \\
         & (0.882) & (1.121) & (0.837) & (1.451)  \\\hline
        SRD: & 0.4862** & 0.6254* & 0.3237 & 0.2436 \\
        $(<-50\%)$ & (0.236) & (0.382) & (0.431) & (0.466) \\\hline
        SRD:& 0.6307*** & 0.7545** & 0.5789** & 0.4886 \\
        $[-50\%,-45\%)$ & (0.233) & (0.332) & (0.292) & (0.846) \\\hline
        SRD:& 0.3642* & 0.1127 & 0.2974 & 0.0262 \\
        $[-45\%,-40\%)$& (0.202) & (0.272) & (0.425) & (0.415) \\\hline
        SRD:& 0.3445 & 0.242 & 0.6138** & 0.0223 \\
        $[-40\%,-35\%)$& (0.233) & (0.290) & (0.279) & (0.432)\\\hline
        SRD:& 0.467*** & -0.0195 & 0.5006* & -0.1933 \\
        $[-35\%,-30\%)$& (0.183) & (0.245) & (0.267) & (0.436)\\\hline
        SRD:& 0.3231* & -0.1767 & 0.4178* & 0.4921 \\
        $[-30\%,-25\%)$& (0.195) & (0.233) & (0.239) & (0.444)\\\hline
        SRD:& 0.1934 & 0.028 & 0.0378 & 0.7571***\\
        $[-25\%,-20\%)$& (0.196) & (0.209) & (0.211) & (0.253)\\\hline
        SRD:& 0.2504 & -0.1155 & 0.0674 & 0.2454\\
        $[-20\%,-15\%)$& (0.200) & (0.198) & (0.218) & (0.269)\\\hline
        SRD:& 0.2193 & -0.1083 & 0.0619 & -0.2322\\
        $[-15\%,-10\%)$& (0.171) & (0.177) & (0.204) & (0.406)\\\hline
        SRD:& 0.0146 & 0.1091 & 0.1712 & -0.1569\\
        $[-10\%,-05\%)$& (0.151) & (0.179) & (0.238) & (0.266)\\\hline
        SRD:& -0.0314 & 0.2073* & -0.1154 & 0.3754*\\
        $[-05\%,0\%)$& (0.150) & (0.124) & (0.256) & (0.202)\\\hline
        SRD:& omitted & omitted & omitted & omitted\\
        $[0\%,05\%)$& & & & \\\hline
        SRD:& 0.0264 & -0.2026 & -0.0804 & -0.0055\\
        $[05\%,10\%)$& (0.1360) & (0.149) & (0.395) & (0.250)\\\hline
        SRD:& -0.0054 & -0.1423 & -0.0226 & -0.0968\\
        $[10\%,15\%)$& (0.165) & (0.148) & (0.246) & (0.247)\\\hline
        SRD:& 0.026 & -0.1789 & 0.1061 & -0.2498\\
        $[15\%,20\%)$& (0.149) & (0.158) & (0.309) & (0.361)\\\hline
        SRD:& -0.152 & -0.1171 & 0.5717** & -0.8346***\\
        $[20\%,25\%)$& (0.297) & (0.182) & (0.267) & (0.340)\\\hline
        SRD:& 0.0913 & -0.2481 & -0.2945 & -0.5186\\
        $[25\%,30\%)$& (0.208) & (0.191) & (0.436) & (0.385)\\\hline
        SRD:& -0.3266 & -0.1904 & 0.7472*** & -0.5058\\
        $[30\%,35\%)$& (0.287) & (0.189) & (0.308) & (0.371)\\\hline
        SRD:& -0.1179 & 0.266 & -0.0413 & -0.5918**\\
        $[35\%,40\%)$& (0.300) & (0.201) & (0.638) & (0.286)\\\hline
        SRD:& -0.2114 & -0.1257 & 1.8206*** & -0.7198**\\
        $[40\%,45\%)$& (0.325) & (0.225) & (0.695) & (0.314)\\\hline
        SRD:& -0.286 & 0.0457 & 1.1126 & -0.6647**\\
        $[45\%,50\%)$& (0.299) & (0.217) & (0.800) & (0.332)\\\hline
        SRD:& 0.0406 & -0.1655 & 0.9579*** & -0.8166***\\
        $[>50\%)$& (0.275) & (0.207) & (0.320) & (0.284)\\\hline
    \end{tabular}
    \smallbreak
    Note: robust standard errors reported in parentheses \\
    significance level: *0.1; **0.05; ***0.01
\end{table}

\pagebreak
\begin{table}[!hbt]
    \centering
    \caption{GEE model results: Lane Distance (measured in Travel Days), asset and non-asset carriers, soft and tight markets}
    \label{tab:Distance}
    \normalsize
    \begin{tabular}{l|c|c|c|c}
        \shortstack{Travel\\ days} & \shortstack{Asset carriers\\ Soft market} & \shortstack{Asset carriers\\ Tight market} & \shortstack{Non-asset carriers\\ Soft market} & \shortstack{Non-asset carriers\\ Tight market}\\\hline\hline
        1 day & -0.0152 & 0.0594 & -0.3804** & 0.3756*\\
         & (0.167) & (0.168) & (0.192) & (0.212) \\\hline
        2 days & omitted & omitted & omitted & omitted \\
         & & & & \\\hline
        3 days & 0.2134 & -0.0518 & 0.0866 & 0.6837***\\
         & (0.220) & (0.162) & (0.268) & (0.206) \\\hline
        4 days & 0.3755 & -0.3351 & -0.2664 & 0.8952**\\
        & (0.260) & (0.244) & (0.363) & (0.409) \\\hline
        $>$4 days & 0.5188 & -0.0202 & 0.3404 & 0.2598\\
         & (0.402) & (0.373) & (0.309) & (0.268)\\\hline
    \end{tabular}
    \smallbreak
    Note: robust standard errors reported in parentheses \\
    significance level: *0.1; **0.05; ***0.01
\end{table}

\begin{table}[!htb]
    \centering
    \caption{GEE model results: Lane Tendering Consistency, asset and non-asset carriers, soft and tight markets}
    \label{tab:Consistency}
    \normalsize
    \begin{tabular}{l|c|c|c|c}
        \shortstack{Consistency\\ variable} & \shortstack{Asset carriers\\ Soft market} & \shortstack{Asset carriers\\ Tight market} & \shortstack{Non-asset carriers\\ Soft market} & \shortstack{Non-asset carriers\\ Tight market}\\\hline\hline
        Cadence: & -0.3690*** & -0.2296** & -0.4965*** & -0.3933\\
        25\% & (0.105) & (0.119) & (0.195) & (0.249) \\\hline
        Cadence: & -0.1291** & -0.1145 & -0.1757 & -0.0943\\
        50\% & (0.063) & (0.088) & (0.141) & (0.152) \\\hline
        Cadence: & omitted & omitted & omitted & omitted \\
        75\% &  & & & \\\hline
        Cadence: & -0.1195 & -0.0611 & 0.243 & 0.199\\
        100\%& (0.093) & (0.077) & (0.198) & (0.200) \\\hline\hline
        Volatility: & 0.9023*** & 0.9203*** & 0.6908*** & 0.6318*\\
        Up to 10\% & (0.243) & (0.219) & (0.169) & (0.373)\\\hline
        Volatility: & 0.4429*** & 0.3591*** & 0.2136 & 0.5434***\\
        (10-25\%] & (0.139) & (0.106) & (0.219) & (0.161)\\\hline
        Volatility: & omitted & omitted & omitted & omitted \\
        (25-50\%] & & & &\\\hline
        Volatility: & -0.3142*** & -0.3789*** & 0.0508 & 0.2485**\\
        (50-75\%] & (0.065) & (0.098) & (0.132) & (0.110)\\\hline
        Volatility: & -0.5076*** & -0.6222*** & -0.050 & 0.1623\\
        (75-100\%] & (0.086) & (0.133) & (0.177) & (0.232)\\\hline
        Volatility: & -0.8049*** & -0.8206*** & -0.1754 & 0.1213\\
        (100-125\%] & (0.119) & (0.133) & (0.254) & (0.210)\\\hline
        Volatility: & -0.7579*** & -0.8419*** & 0.0278 & 0.1355\\
        (125-150\%] & (0.116) & (0.140) & (0.285) & (0.220)\\\hline
        Volatility: & -0.6011*** & -0.8065*** & -0.0155 & 0.1349\\
        (150-200\%] & (0.127) & (0.182) & (0.304) & (0.238)\\\hline
        Volatility: & -0.4359*** & -0.746*** & 0.1467 & 0.2419\\
        Over 200\% & (0.143) & (0.175) & (0.313) & (0.266)\\\hline
    \end{tabular}
    \smallbreak
    Note: robust standard errors reported in parentheses \\
    significance level: *0.1; **0.05; ***0.01
\end{table}

\begin{table}[!htb]
    \centering
    \caption{GEE model results: Surge Volume, asset and non-asset carriers, soft and tight markets}
    \label{tab:Surge}
    \normalsize
    \begin{tabular}{l|c|c|c|c}
        \shortstack{Surge \\ Category} & \shortstack{Asset carriers\\ Soft market} & \shortstack{Asset carriers\\ Tight market} & \shortstack{Non-asset carriers\\ Soft market} & \shortstack{Non-asset carriers\\ Tight market}\\\hline\hline
        Within & 0.1757** & 0.1009 & 0.0181 & 0.2668*\\
         Mean & (0.089) & (0.084) & (0.251) & (0.146) \\\hline
        Mean to & omitted & omitted & omitted & omitted \\
        10\% Surge & & & & \\\hline
        10-20\% & -0.0203 & -0.1837** & 0.1271 & 0.124\\
        Surge & (0.108) & (0.094) & (0.294) & (0.150)\\\hline
        Over 20\% & -0.1559** & -0.2334*** & -0.1387 & -0.0099\\
        Surge& (0.079) & (0.083) & (0.265) & (0.120)\\\hline
    \end{tabular}
    \smallbreak
    Note: robust standard errors reported in parentheses \\
    significance level: *0.1; **0.05; ***0.01
\end{table}

% \begin{table}[!hbt]
%     \centering
%     \caption{GEE model results: Load Tender Lead Time, asset and non-asset carriers, soft and tight markets}
%     \label{tab:TLT}
%     \footnotesize
%     \begin{tabular}{l|c|c|c|c}
%         \shortstack{Tender Lead \\ Time} & \shortstack{Asset carriers\\ Soft market} & \shortstack{Asset carriers\\ Tight market} & \shortstack{Non-asset carriers\\ Soft market} & \shortstack{Non-asset carriers\\ Tight market}\\\hline\hline
%         1 day & 0.3869*** & 0.725*** & 0.5573*** & 0.6794***\\
%          & (0.088) & (0.102) & (0.161) & (0.174)\\\hline
%         2 days & -0.1404** & -0.1327** & -0.0994 & 0.1302\\
%          & (0.062) & (0.065) & (0.145) & (0.107)\\\hline
%         3 days & omitted & omitted & omitted & omitted \\
%          & & & & \\\hline
%         4 days & -0.1353** & -0.0445 & -0.2563*** & -0.0195\\
%          & (0.063) & (0.065) & (0.097) & (0.114)\\\hline
%         5 days & -0.0168 & -0.1234 & -0.0846 & 0.1457\\
%          & (0.078) & (0.082) & (0.151) & (0.187)\\\hline
%         6 days & 0.0127 & -0.0661 & -0.2691** & 0.2249\\
%          & (0.086) & (0.089) & (0.124) & (0.169)\\\hline
%         7 days & 0.0246 & -0.041 & -0.2119 & 0.0526\\
%          & (0.100) & (0.089) & (0.184) & (0.202)\\\hline
%         >7 days & -0.4744*** & -0.503*** & -0.8722*** & -0.3008\\
%          & (0.139) & (0.109) & (0.188) & (0.265)\\\hline
%     \end{tabular}
%     \smallbreak
%     Note: robust standard errors reported in parentheses \\
%     significance level: *0.1; **0.05; ***0.01
% \end{table}

\begin{table}[!htb]
    \centering
    \caption{GEE model results: Carrier Fleet Size, asset carriers, soft and tight markets}
    \label{tab:fleet}
    \normalsize
    \begin{tabular}{l|c|c}
        \shortstack{Fleet size\\ No. tractors} & \shortstack{Asset carriers\\ Soft market} & \shortstack{Asset carriers\\ Tight market}\\\hline\hline
        Log Tractor Count & -0.0109 & -0.0088 \\
         & (0.079) & (0.103)  \\\hline
    \end{tabular}
    \smallbreak
    Note: robust standard errors reported in parentheses \\
    significance level: *0.1; **0.05; ***0.01
\end{table}

\begin{table}[!htb]
    \centering
    \caption{GEE model results: Shipper Fixed Effects, asset and non-asset carriers, soft and tight markets}
    \label{tab:shipperfixed}
    \normalsize
    \begin{tabular}{l|c|c|c|c}
        \shortstack{Shipper fixed\\ effects variable} & \shortstack{Asset carriers\\ Soft market} & \shortstack{Asset carriers\\ Tight market} & \shortstack{Non-asset carriers\\ Soft market} & \shortstack{Non-asset carriers\\ Tight market}\\\hline\hline
        Shipper size & 0.4033** & 0.284 & 0.5908** & 0.3054\\
        (log monthly volume) & (0.177) & (0.271) & (0.27) & (0.406)\\\hline
        Vertical: & 0.655*** & 0.8121*** & -0.3843 & 0.2414\\
        Automotive & (0.216) & (0.261) & (0.279) & (0.451)\\\hline
        Vertical: & 0.9596*** & 0.9327*** & -0.4351 & 0.1938\\
        F\&B/CPG & (0.194) & (0.280) & (0.266) & (0.439)\\\hline
        Vertical: & omitted & omitted & omitted & omitted \\
        Paper \& Packaging & & & &  \\\hline
        Vertical: & 0.5557 & 0.5942* & -0.8417* & 0.5001\\
        Manufacturing & (0.381) & (0.336) & (0.468) & (0.531)\\\hline
        Vertical: & -0.2443 & 0.9402 & 1.3812*** & -0.2445\\
        Other & (0.602) & (0.636) & (0.467) & (0.684)\\\hline
    \end{tabular}
    \smallbreak
    Note: robust standard errors reported in parentheses \\
    significance level: *0.1; **0.05; ***0.01
\end{table}

% \end{appendices}

\pagebreak
\end{document}